\documentclass[11pt,a4paper]{article}
\usepackage{epsfig,axodraw}
\usepackage{array,cite,amsmath,amssymb}
\usepackage[flushleft]{threeparttable}
\usepackage{subcaption}
\usepackage{kantlipsum}
\usepackage{float}
\usepackage[usenames,dvipsnames]{xcolor}
\usepackage[breakable, theorems, skins]{tcolorbox}
\tcbset{enhanced}
\usepackage{booktabs}
\usepackage{caption}
\usepackage{threeparttable}
\usepackage{multirow}
\DeclareRobustCommand{\mybox}[2][gray!20]{%
\begin{tcolorbox}[   %% Adjust the following parameters at will.
        breakable,
        left=0pt,
        right=0pt,
        top=0pt,
        bottom=0pt,
        colback=#1,
        colframe=#1,
        width=\dimexpr\textwidth\relax, 
        enlarge left by=0mm,
        boxsep=5pt,
        arc=0pt,outer arc=0pt,
        ]
        #2
\end{tcolorbox}
}
\newcommand{\newc}{\newcommand}
\newc{\gev}{\,GeV}
\newcolumntype{M}[1]{>{\centering\arraybackslash}m{#1}}
\newcolumntype{N}{@{}m{0pt}@{}}
\newc{\mev}{\,MeV}
\newc{\ra}{\rightarrow}
\newc{\rpv}{$\mathrm{\not\!R_p}$}
\newc{\rp}{$\mathrm{R_p}$}
\newc{\real}{\mathcal{R}e}
\newc{\alsm}{{\displaystyle \sum_{\alpha=1,2}}}
\newc{\besm}{{\displaystyle \sum_{\beta=1,2}}}
\newc{\al}{\alpha}
\newc{\sgn}{\mr{sgn}\,}
\newc{\be}{\beta}
\newc{\ga}{\gamma}
\newc{\de}{\delta}
\newc{\sla}{\!\!\!\!\!\not\:\:\!}
\newc{\slab}{\!\!\!\!\!\not\,\,\,}
\newc{\slac}{\!\!\!\!\!\!\!\not\,\,\,\,}
\newc{\met}{$\not\!\!E_T$}
\newc{\cw}{\cos\theta_W}
\newc{\sw}{\sin\theta_W}
\newc{\ssw}{\sin^2\theta_W}
\newc{\ccw}{\cos^2\theta_W}
\newc{\cbe}{\cos\beta}
\newc{\sbe}{\sin\beta}
\newc{\ort}{\frac1{\sqrt{2}}}
\newc{\sh}{\hat{s}}
\newc{\uh}{\hat{u}}
\newc{\tha}{\hat{t}}
\newc{\sa}{\sin\al}
\newc{\ca}{\cos\al}
\newc{\mz}{M_{\mr{Z}}}
\newc{\mw}{M_{\mr{W}}}
\newc{\bv}{$\mathrm{\not\!B}$}
\newc{\lv}{$\mathrm{\not\!L}$}
\newc{\beq}{\begin{equation}}
\newc{\eeq}{\end{equation}}
\newc{\ie}{{\it i.e.\/}\ }
\newc{\lam}{\lambda}
\newc{\cht}{\tilde{\chi}}
\newc{\glt}{\tilde{g}}
\newc{\upt}{\tilde{u}}
\newc{\qkt}{\tilde{q}}
\newc{\elt}{\tilde{\ell}}
\newc{\hgt}{\tilde{H}}
\newc{\nut}{\tilde{\nu}}
\newc{\dnt}{\tilde{d}}
\newc{\ftl}{\mr{\tilde{f}}}
\newc{\psb}{\bar{\psi}}
\newc{\rtt}{2^{1/2}}
\newc{\mut}{\tilde{\mu}}
\newc{\mr}{\mathrm}
\newc{\bath}{\bar{\theta}}
\newc{\tht}{\theta}
\newc{\JC}{{\bf J}}
\newc{\lra}{\longrightarrow}
\newc{\eg}{{\it e.g.\  }}
\newc{\barr}{\begin{eqnarray}}
\newc{\earr}{\end{eqnarray}}
\newc{\me}{\mathcal{M}}
\newc{\dbm}{\partial_\mu}
\newc{\dbmu}{\stackrel{\leftrightarrow\  }{\partial^\mu}}
\newc{\sgm}{\sigma_\mu}
\newc{\captionB}[2]{\caption[{#1}]{{\small {#2}}}}
\newc{\ahref}[2]{#2}
\pdfoutput=1 

\usepackage{jheppub} 
\title{\boldmath Hadronization and Decay of Excited Heavy Hadrons in Herwig~7}

\author{M.R.~Masouminia,}
\author{P.~Richardson}

\affiliation{Institute for Particle Physics Phenomenology, Durham University, Durham, UK}

\emailAdd{mohammad.r.masouminia@durham.ac.uk}
\emailAdd{peter.richardson@durham.ac.uk}

\abstract{ \small
We revisit the hadronization and decay of excited heavy mesons and heavy baryons in \textsf{Herwig~7} general-purpose event-generator, following four distinct steps: (i) Passing through the polarisation of heavy hadrons at the end of parton shower through the application of heavy quark effective theory (HQET), where the emergence of a spin-flavour symmetry allows for the determination of the polarisations of the excited heavy mesons and heavy baryons from the helicity states of the light and heavy quarks. (ii) Improving the strong and radiative decay modes of the excited heavy mesons, where in the absence of conclusive experimental data on many of the decays, one needs to rely on HQET symmetries to determine the favoured decay modes, widths and branching ratios. (iii) Re-examination of the production rates of heavy hadrons using \textit{all} available experimental data sources and (iv) performing a general tune for \textsf{Herwig}'s free parameters to reflect the implemented changes. We compare our predictions against existing experimental data in the presence/absence of the newly implemented updates. These improvements will be available with \textsf{Herwig-7.3.0} public release.
}

\begin{document}
\noindent{\hfill \small IPPP/23/52 \\[0.1in]}
\maketitle
\flushbottom

\section{Introduction}
\label{sec:intro}

% about the status of Herwig
In the study of high-energy particle physics, \textsf{Herwig~7} general-purpose event generator stands as a powerful tool, providing a platform to simulate the intricate processes of hadron formation and decay~\cite{Bahr:2008pv,Bellm:2015jjp,Bellm:2017bvx,Bellm:2019zci}. One of the natural steps in the simulation of high-energy collisions in \textsf{Herwig~7} is the emergence of observable hadronic states through fragmentation and decay of quarks and gluons at the end of a sequential parton shower via an overall non-perturbative cluster hadronization model~\cite{Kupco:1998fx}. In this context, it would be productive to transmit the polarisation information of the heavy quarks appearing at the end of the parton shower to the hadronization handler to accurately account for the net polarisation observed in excited mesons and heavy baryons. This information ensures that the interplay between quark spins and flavours is appropriately considered during the hadronization process, leading to a more realistic depiction of the polarisation properties of the resulting particles. This is possible by implementing the effects of the Heavy Quark Effective Theory (HQET)~\cite{Falk:1990yz,Neubert:1993mb,Falk:1993dh,Bigi:1994em}.

% HQET and spin-symmetry
HQET is a framework that provides a systematic way to describe the behaviour and interactions of heavy quarks within hadrons. In this context, a quark is considered ``heavy" if its mass is much larger than the QCD energy scale, $\Lambda_{\rm QCD}$. This heavy quark is then treated as moving at a sub-relativistic velocity. HQET takes advantage of this mass hierarchy and velocity separation to simplify the mathematical description of heavy quark interactions within hadrons, assuming that the velocity and the spin of the heavy quarks can be treated as independent variables, resulting in a decoupling between the dynamics of the heavy quark and the lighter degrees of freedom. This allows for a separation of scales, where the interactions involving the heavy quark can be analysed using a perturbative expansion while the interactions involving the lighter degrees of freedom are treated non-perturbatively. The emergence of this \textit{heavy quark spin-flavour symmetry} and the associated effects allow for model-independent predictions of certain properties of heavy hadrons, such as their spectra and decay rates.

% Why revisit decays
The above endeavour involves modelling the decays of excited heavy hadrons, specifically those containing a single charm or bottom quark. In the framework of heavy quark symmetry, these decays are not only determined by the possibilities of available decay modes but also exhibit inherent relationships between the associated couplings. The limited experimental data concerning many of these decays necessitates a reliance on these symmetries to determine the decay modes, widths and branching ratios. Therefore, it would be pertinent to refine the existing prescriptions for the electromagnetic and strong-isospin-violating decay modes of excited heavy mesons while employing the effects of HQET. The latter modes become more significant in cases where strong isospin-conserving decays are either disallowed or significantly kinematically suppressed.

% Why revisit rates and why tuned
Undoubtedly, incorporating the heavy quark spin-flavour symmetry and consequent enhancements to the heavy excited hadronic states, particularly the isospin-violating decay modes, will trigger a cascade of effects that can reverberate through \textsf{Herwig}'s hadronization and decay handlers, thereby influencing the production rates of all types of hadronic states. Furthermore, the introduction of new model-independent parameters into \textsf{Herwig}'s cluster hadronization model, with the aim of refining the kinematic mass threshold that governs cluster splittings holds the potential to ripple across both baryon and meson production rates. This prompts the necessity of undertaking a general tuning for hadronization parameters, alongside the pertinent parton shower parameters.

% outlook
In this study, we outline the advancements made in predictive modelling of heavy hadron production, focusing on enhancing \textsf{Herwig~7} event generator. Section~\ref{sec:HQET} provides a thorough examination of HQET and Spin-Flavour Symmetry, breaking it down further into the kinematics of fragmentation, and the polarisation mechanisms for excited heavy mesons and baryons. Section~\ref{sec:decay} discusses the decay processes of these excited heavy mesons, laying the groundwork for Section~\ref{sec:hadronresults}, which scrutinises relevant observables. In Section~\ref{sec:tune}, a comprehensive tuning methodology for \textsf{Herwig-7.3.0} is introduced, driven by a brute-force, multilayered approach to parameter optimisation. Section~\ref{sec:rate} then evaluates the efficacy of these modifications by analysing heavy hadron production rates using an extensive dataset. A brief summary and conclusions are presented in Section~\ref{sec:conc}. The paper also includes an appendix detailing the signature of heavy hadrons in the $\Upsilon(4S)$ continuum region, providing further context for the validation of our study.

% how calculations are done
The calculations of induced polarisations of excited heavy mesons and heavy baryons closely follow the methodology of~\cite{Falk:1993rf}, while the enhancements to decay modes build upon the notation of~\cite{Falk:1992cx}. The results showcased herein are generated using \textsf{Herwig-7.3.0}, incorporating internally generated matrix elements and \textsf{Herwig}'s \texttt{QCD$\oplus$QED$\oplus$EW} parton shower~\cite{Masouminia:2021kne,Darvishi:2021het}. The subsequent analysis is conducted employing \textsf{Rivet} (v3.1.8)~\cite{Buckley:2010ar}. Our tuning endeavours draw inspiration from the methodology outlined in \textsf{Professor2} (v2.3.7)~\cite{Buckley:2009bj}, albeit substantially adapted to accommodate the tuning of $12$ independent parameters. The outcomes of this comprehensive exploration, embedded within the \textsf{Herwig~7} general-purpose event generator, will be made publicly accessible alongside the anticipated release of \textsf{Herwig-7.3.0}.

\section{Heavy Quark Effective Theory and Spin-Flavour Symmetry}
\label{sec:HQET}

A hadronic state can be considered as a non-recoiling source of colour if its constituents can be divided into separate heavy ($Q$) and light ($q$) degrees of freedom, while $m_Q \gg \Lambda_{\rm QCD}$~\cite{Shuryak:1980pg, Caswell:1985ui, Eichten:1987xu, Lepage:1987gg, Isgur:1989vq}. In such a scenario, the effects associated with the heavy quark colour magnetic moment decouple from hadronic properties, leading to the emergence of heavy quark spin-flavour symmetry~\cite{Falk:1993rf}. This symmetry can be used to produce model-independent predictions of heavy hadron spectra, weak matrix elements, and strong decay rates. An interesting side-effect of this spin-flavour symmetry is the ability to distinguish between short- and long-distance interactions in a hadronization sequence, such as at the end of a procedurally generated parton shower.

Strongly interacting particles produced during hard processes and subsequent parton showers in a simulated event generally have relatively large momenta. The corresponding perturbative interactions during these stages occur on short time scales. In contrast, the non-perturbative fragmentation processes that follow are inherently long-distance and thus occur over longer time scales. The ensuing hadronization processes, regardless of their model dependencies, are expected to take place at length scales of the order of $\mathcal{O}(\Lambda_{\rm QCD}^{-1})$. This results in energy redistributions much smaller than $m_Q$, making it plausible to assume that the velocity, mass, and spin of the heavy quark state (determined by the calculable short-distance physics) remain unchanged, suggesting a decoupling between the dynamics of the heavy and light quark states~\cite{Falk:1990cz,Mannel:1990up,Cohen:1992hu}. 

This argument extends to the production of excited heavy mesons and baryons in a hadronization event, provided the produced hadrons can be tagged with the helicity information of their heavy quark constituents~\cite{Falk:1993rf}. This tagging usually comes with two conditions: (i) If the phase-space of the light-quark state allows its angular momentum to become of the order of $\mathcal{O}(m_Q^{-1})$, the process can redistribute this angular momentum, meaning the outgoing heavy quark polarisation will depend on the polarisation of the light degree of freedom created in the fragmentation process. (ii) While conserving parity, fragmentation can produce anisotropic light degrees of freedom along its axis. The alignment of a light degree of freedom with spin $j$ may be characterised by a model-dependent dimensionless parameter, $\omega_j$~\cite{Falk:1993rf}.

Additionally, we detail the interactions for charm mesons, though the derived results are also applicable to bottom mesons. Focusing initially on the $s$- and $p$-wave mesons, we identify three meson multiplets under the auspices of heavy quark symmetry, as presented in \cite{Falk:1992cx}:
\begin{itemize}
\item[(i)] The $J^P=0^-, 1^-$ doublet (the ground state),
\item[(ii)] The $J^P=1^+, 2^+$ doublet,
\item[(iii)] The $J^P=0^+, 1^+$ doublet.
\end{itemize}
Furthermore, one must consider the potential for mixing between heavy mesons, which arises due to sub-leading corrections within the heavy quark limit.

One should note that in this study, helicity is used to describe the spin polarisation of heavy quarks. Although not a perfect quantum number for massive particles, helicity is a practical and stable descriptor in the non-relativistic regime of HQET. In this framework, the heavy quark's large mass results in a decoupling of its spin and velocity from the lighter degrees of freedom, allowing helicity to remain stable over the time scales relevant to hadronization and decay processes. This stability ensures that the helicity of the heavy quark can significantly influence the polarisation of the resulting hadrons, as detailed in Ref.~\cite{Falk:1993rf}.

\subsection{Kinematics of Fragmentation}

To properly translate the implications of HQET and the above-mentioned spin-flavour symmetry, one has to start by projecting the kinematics and the time scales of heavy quark fragmentation. Here we closely follow the footsteps of~\cite{Falk:1993rf}, by assuming to work in the rest frame of the heavy quark while identifying the preferred direction as the momentum of the parton shower progenitor. In this setup, the only interaction that can manipulate the spin of the heavy quark, i.e. the colour magnetic moment of the light degrees of freedom, is sufficiently suppressed by a factor of $m_Q^{-1}$. Collectively, the hadron can possess a total spin of $j_{\pm} = j_q \pm 1/2$, with $j_q$ being the spin of the light state. We can, therefore, identify the $j_{\pm}$ configurations with a spin multiplet ($H$, $H^\star$), and assume a small mass splitting $\Delta m = m_{H^\star} - m_{H}$. 

Because they contain a heavy quark, the $H$ and $H^\star$ states are unstable. Hence, they can decay weakly to lighter hadrons, where it is safe to assume an identical decay rate 
\begin{equation}
\Gamma(H \to X) = \Gamma(H^\star \to X).
\end{equation}
Alternatively, $H^\star$ can strongly or radiatively decay to $H$, with rate 
\begin{equation}
\gamma(H^\star \to H X) \propto \Phi_{\rm phase-space}^{H^\star \to H X} \times |\mathcal{M}(H^\star \to H X)|^2 \sim \mathcal{O}(m_q^{-(2+n)}),
\label{gamma}
\end{equation}
where $n \geq 1$, suggesting $\Delta m \gg \gamma$. Meanwhile, the relation between $\Gamma$, $\Delta m$ and $\gamma$ would determine the nature of the fragmentation. Particularly, the case of $\Gamma \gg \Delta m \gg \gamma$ would allow for rapid decay of heavy hadrons while decoupling the colour magnetic moments of the heavy and the light degrees of freedom. This would be valid for both strong or weak decays and even for $\Gamma \sim \Lambda_{\rm QCD}$, where the heavy quark can go through partial hadronization before it decays. Both other possible cases, namely $\Delta m \gg \Gamma \gg \gamma$ and $\Delta m \gg \gamma \gg \Gamma$ result in depolarisation of the heavy quark from its initial orientation. The above argument suggests that under most conditions, the angular distribution of decay products gives no information on the polarisation of the heavy quark, unless the condition $\Gamma \gg \Delta m \gg \gamma$ is realised. This is the so-called Falk-Peskin ``no-win'' theorem, suggesting that the polarisation of heavy quarks can remain intact only though strong, weak or radiative decays of heavy excited mesons and heavy baryons~\cite{Falk:1993rf}. 

\subsection{polarisation of Excited Heavy Mesons}
\label{Mesons}

To explore the ramifications of HQET and spin-flavour symmetry regarding the polarisations of heavy excited mesons, we initially focus on the charm sector, specifically the observed excited charmed mesons $D_0^{\star}$, $D_1$, $D_1'$ and $D_2^{\star}$. This discussion can be readily extended to bottom mesons. In this framework, the spins of the light and heavy quarks in the first and second excited $D$ mesons combine to form the total angular momentum $j$. For the first excited $D$ mesons, a spin-${1 \over 2}$ light quark couples with a spin-${1 \over 2}$ heavy quark, resulting in a total angular momentum $j = 1$. This combination leads to the quantum numbers $j^P = 0^+$ or $j^P = 1^+$, corresponding to the $D_0^{\star}$ and $D_1$ states, respectively\footnote{At order $1/m_c$, there is mixing between the $D_1$ and the $D_1'$ states due to their identical quantum numbers.}. For the second excited $D$ mesons, the light and heavy quarks possess higher orbital angular momentum, resulting in quantum numbers $j^P = \left(3 \over 2\right)^+$, which corresponds to the $D_2^{\star}$ charmed state.

The colour magnetic interaction, due to its associated time scale, becomes effectively decoupled from the dynamics of the heavy quark spin. This decoupling occurs because the colour magnetic interaction operates over a short time scale compared to the hadronization process. As a result, the spin orientation of the light quark becomes independent of the heavy quark’s spin during the formation of the excited mesons. This allows the initial polarisation of the heavy quark to influence the polarisation of the resulting hadron, as the heavy quark's spin information is preserved through the hadronization process.

Assuming that the initial $c$ quark entering the fragmentation processes has left-handed helicity, $j_Q = -1/2$, a heavy charmed meson can be formed by combining with light degrees of freedom with $j_q = {1/2}$. One would expect that the colour magnetic interaction becomes decoupled from in such a process, due to its time scale, leaving the spin-orientation of the light degrees independent of the charm quark and distributed uniformly, i.e. $j_q^{(3)} = \pm 1 / 2$ with equal probabilities:
\begin{equation}
{\left|\downarrow\,\rangle \right.}_c \; {\left|\downarrow\,\rangle \right.}_q
\quad {\rm and } \quad
{\left|\downarrow\,\rangle \right.}_c \; {\left| \uparrow\,\rangle \right.}_q 
\nonumber
\end{equation}
This scenario becomes more intriguing when considering the $D_1$ and $D_2^{\star}$ meson states. In conjunction with the assumed left-handed helicity of the charm quark, we now have light degrees of freedom with $j_q = 3/2$, which can manifest in any of the four possible helicity states. Parity invariance dictates that the probability of forming a specific helicity state cannot rely on the sign of its helicity, $j_3$, although probabilities may differ for states with distinct helicity magnitudes, $|j_q^{(3)}|$. Defining the parameter $\omega_j$ ($0 \leq \omega_j \leq 1$), that is the likelihood of fragmentation leading to a state with the maximum value of $|j_q^{(3)}|$ in a system with light degrees of freedom with spin $j_q$, one may derive the probabilities for the emergence of different helicity states of the light degrees as
\begin{table}[H]
\centering
\resizebox{0.5\textwidth}{!}{%
\begin{tabular}{|c||c c c c|}
\hline
$j_q^{(3)}$ & -3/2 & -1/2 & 1/2 & 3/2 \\
\hline \hline
Probability & ${1 \over 2} \omega_{3 \over 2}$ 
            & ${1 \over 2} (1-\omega_{3 \over 2})$ 
            & ${1 \over 2} (1-\omega_{3 \over 2})$ 
            & $\begin{matrix}  \\ \\ \end{matrix} {1 \over 2} \omega_{3 \over 2} \begin{matrix}  \\ \\ \end{matrix}$
\\ \hline
\end{tabular}}
\caption{Probability distribution of the helicity of the light degrees in excited heavy mesons.}
\end{table}
{\noindent The combination of the left-handed $c$ helicity state with a specific light degrees of freedom helicity $j_q^{(3)}$ results in a coherent linear superposition of the charmed states with helicity $j = j_q + j_Q$. This rationale leads to the following populated helicity states and their corresponding probabilities for $D$, $D_0^{\star}$, $D_1$ and $D_2^{\star}$~\cite{Falk:1993rf}:}
\begin{table}[H]
\centering
\resizebox{0.5\textwidth}{!}{%
\begin{tabular}{|c||c c c c c|}
\hline
$j^{(3)}$ & $-2$ & $-1$ & $0$ & $+1$ & \hspace{0.1in} $\begin{matrix}  \\ \\ \end{matrix} +2 \begin{matrix}  \\ \\ \end{matrix}$ \hspace{0.1in} 
\\ \hline \hline
$D$ & $-$ & $-$ & ${1 \over 4}$ & $-$ & $\begin{matrix}  \\ \\ \end{matrix} - \begin{matrix}  \\ \\ \end{matrix}$ 
\\
$D^{\star}$ & $-$ 
            & ${1\over 2}$ 
            & ${1\over 4}$ 
            & $0$ 
            & $\begin{matrix}  \\ \\ \end{matrix} - \begin{matrix}  \\ \\ \end{matrix}$ 
\\ 
$D_1$ & $-$ 
      & ${1 \over 8} (1-\omega_{3 \over 2})$
      & ${1 \over 4} (1-\omega_{3 \over 2})$
      & ${3 \over 8}(1-\omega_{3 \over 2})$
      & $\begin{matrix} \\ \\ \end{matrix}-\begin{matrix}  \\ \\ \end{matrix}$ 
\\ 
$D_2^{\star}$ & ${1 \over 2} \omega_{3 \over 2}$
      & ${3 \over 8}(1-\omega_{3 \over 2})$
      & ${1 \over 4} (1-\omega_{3 \over 2})$
      & ${1 \over 8} \omega_{3 \over 2}$
      & $\begin{matrix} \\ \\ \end{matrix} 0 \begin{matrix} \\ \\ \end{matrix} $
\\ \hline
\end{tabular}}
\caption{Probabilities for the population of the possible helicity states of $D$, $D^{\star}$, $D_1$, and $D_2^{\star}$ mesons, for the specific case of the initial heavy quark having negative helicity.}
\label{tab2}
\end{table}
{\noindent Note that since $j^{(3)} = +1$ ($j^{(3)} = +2$) is only conceivable for the $D^{\star}$ ($D_2^{\star}$) state, whereas based on the initial assumption $j_Q^{(3)} = -1/2$, its corresponding probability would be zero.}

To evaluate the parameter $\omega_j$, we consider the amplitude for the production of a pion at $\theta,\phi$ from a $H^{\star} \to H\pi$ type meson decay, which is proportional to the spherical harmonics $Y_{j}^{\ell}(\theta, \phi)$ ($\ell$ being the angular momentum quantum number of $H^{\star}$)
\begin{equation}
{d\Gamma(H^{\star} \to H\pi) \over d\cos\theta} \propto \int d\phi \sum_{j}  P_{H^{\star}}(j)  \bigl| Y_{j}^{\ell}(\theta,\phi) \bigr|^2,
\label{diff_width}
\end{equation} 
where $P_{H^{\star}}(j)$ are the probabilities given in Table~\ref{tab2}. Additionally, $\theta$ and $\phi$ are the angles of emission for the produced pion. Considering the case of $D_2^{\star} \to D\pi$, Eq.~\eqref{diff_width} can be rewritten as
\begin{equation}
{1 \over \Gamma}{d\Gamma(D_2^{\star} \to D\pi) \over d\cos\theta} = {1 \over 4} 
\left[ 1 + 3 \cos^2 \theta - 6 \; \omega_{3 \over 2} \left( \cos^2 \theta - {1 \over 3} \right)\right].
\label{diff_width_10}
\end{equation} 
In Ref.~\cite{Falk:1993rf}, a comprehensive analysis over differential decay width Eq.~\eqref{diff_width_10} and experimental data from~\cite{Isgur:1991wq,ARGUS:1988ujx,ARGUS:1989mcc,CLEO:1989qui,Lu:1991px} resulted in predicting $\omega_{3 \over 2} < 0.24$ with in 90\% CL. 

To implement the above arguments in the context of HQET and spin-flavour symmetry principles in \textsf{Herwig~7}, firstly we generalise Table~\ref{tab2} for an arbitrary heavy quark helicity, $\rho_Q$:
\begin{table}[H]
\centering
\resizebox{\textwidth}{!}{%
\begin{tabular}{|c||c c c c c|}
\hline
$\hat{\rho}$ & $\rho_{0,0}$ & $\rho_{1,1}$ & $\rho_{2,2}$ & $\rho_{3,3}$ & $\rho_{4,4}$ 
\\ \hline \hline
$D$ & $1$ & $-$ & $-$ & $-$ & $\begin{matrix}  \\ \\ \end{matrix} -\begin{matrix}  \\ \\ \end{matrix}$ 
\\
$D^{\star}$ & ${1\over 2}(1-\rho_Q)$
      & ${1\over 2}$
      & ${1\over 2}(1+\rho_Q)$
      & $-$
      & $\begin{matrix}  \\ \\ \end{matrix} -\begin{matrix}  \\ \\ \end{matrix}$ 
\\ 
$D_1$ & ${1\over 16}[1-\rho_Q + \omega_{3 \over 2}(3-5\rho_Q)]$
      & ${1 \over 4}(1-\omega_{3 \over 2})$
      & ${1\over 16}[1-\rho_Q + \omega_{3 \over 2}(3+5\rho_Q)]$
      & $-$
      & $\begin{matrix} \\ \\ \end{matrix}-\begin{matrix}  \\ \\ \end{matrix}$ 
\\ 
$D_2^{\star}$ & ${1 \over 4} \omega_{3 \over 2} (1-\rho_Q)$
      & ${3 \over 16}(1-\rho_Q) - {1 \over 8}\omega_{3 \over 2}(1-\rho_Q) $
      & ${1 \over 4} (1-\omega_{3 \over 2})$
      & ${3 \over 16}(1+\rho_Q) - {1 \over 8}\omega_{3 \over 2}(1+\rho_Q) $
      & $\begin{matrix} \\ \\ \end{matrix} 
        {1 \over 4} \omega_{3 \over 2} (1+\rho_Q)
        \begin{matrix} \\ \\ \end{matrix} $
\\ \hline
\end{tabular}}
\caption{Possible polarisation states of charmed mesons, $D$, $D^{\star}$, $D_1$ and $D_2^{\star}$.}
\label{tab3}
\end{table}
{\noindent Here, $\hat{\rho}$ is the diagonal polarisation matrix of the produced mesons, with its components $\rho_{i,i}$, $i=0,1,2,3,4$ running through the most negative to most positive valid helicity states.} The value $\omega_{3 \over 2} = 0.20$ is assumed as default. 

In \textsf{Herwig~7}, the above implementation is remitted through the introduction of a so-called \texttt{SpinHadronizer} transitional class. In particular, its \texttt{mesonSpin} sub-module systematically handles the assignment of spin information and polarisations to generated mesons based on their spin characteristics and heavy constituent quark flavours. Starting with checks on the meson's parentage and its constituents, \texttt{mesonSpin} employs spin information from the heavy quark to construct the spin properties of the meson, pertinent to the excited heavy mesons (in addition to their complex conjugates):
\begin{itemize}

\item[(i)] $D^{\star +}$, $D^{\star 0}$, $D_s^{\star +}$, $B^{\star 0}$, $B^{\star +}$, $B_s^{\star 0}$

\item[(ii)] $K_1^0$, $K_1^+$, $h'_1$, $D_1^+$, $D_1^0$, $D_{s1}^+$, $D_1^{' +}$, $D_1^{' 0}$, $D_{s1}^{' +}$, $B_1^0$, $B_1^+$, $B_{s1}^0$

\item[(iii)] $K^{\star 0}_2$, $K^{\star +}_2$, $f'_2$, $D^{\star +}_2$, $D^{\star 0}_2$, $D_{s2}^+$, $B_2^0$, $B_2^+$, $B_{s2}^0$

\end{itemize}
The above items correspond to the 2nd, 3rd and 4th row of Table~\ref{tab3}, respectively. the \texttt{mesonSpin} sub-module also calculates the helicity of the quark, updates the average polarisation for the specific flavour and then assigns spin density matrix elements to the meson according to its spin value. It accommodates different spin types and meson categories, tailoring the spin behaviour and polarisation calculations to match the distinct characteristics of various excited heavy mesons.

\subsection{polarisation of Heavy Baryons}
\label{Baryons}

The ground state of a heavy baryon is characterised by a heavy quark combined with a $j_{qq} = 0$ helicity arrangement of a diquarks system. In this configuration, there is no angular momentum available to be transferred to the heavy quark, resulting in the preservation of the initial polarisation without dilution. This implies that the polarisation of the initial heavy quarks can indeed influence the ground state of heavy baryons. Here, the relative probabilities of finding these states in heavy sector fragmentation are still governed by two parameters, $\omega_a$ and $\omega_j$. The parameter $\omega_a$ represents the relative probability of producing a $j_{qq} = 1$ diquark as opposed to the ground state $j_{qq} = 0$ configuration. In the context of \textsf{Hrewig~7}'s cluster hadronization model, we take $\omega_a = 1$ as in general no preference exists between spin-0 and spin-1 diquarks.

Now, similar to the case of excited heavy mesons, we consider the fragmentation of a $c$ quark with an assumed left-handed polarisation. Given values of $\omega_a$ and $w_1$, the various helicity states of the charmed baryons can be populated by fragmentation
according to the following table~\cite{Falk:1993rf}:
\begin{table}[H]
\centering
\resizebox{0.45\textwidth}{!}{%
\begin{tabular}{|c||c c c c|}
\hline
$j^{(3)}$ & $-3/2$ 
          & $-1/2$ 
          & $+1/2$ 
          & $\begin{matrix}  \\ \\ \end{matrix} +3/2 \begin{matrix}  \\ \\ \end{matrix}$ 
\\ \hline \hline
$\Lambda_c$ & $-$ 
            & ${1\over 1+\omega_a}$ 
            & $0$ 
            & $\begin{matrix}  \\ \\ \end{matrix}-\begin{matrix}  \\ \\ \end{matrix}$ 
\\ 
$\Sigma_c$ & $-$
           & ${(1-\omega_1) \omega_a \over 3 (1+\omega_a)}$
           & ${ \omega_1 \omega_a \over 3 (1+\omega_a)}$
           & $\begin{matrix} \\ \\ \end{matrix}-\begin{matrix}  \\ \\ \end{matrix}$ 
\\ 
$\Sigma_c^\star$ & ${\omega_1 \omega_a \over 2 (1+\omega_a)}$
      & ${2 (1-\omega_1) \omega_a \over 3 (1+\omega_a)}$
      & ${\omega_1 \omega_a \over 6 (1+\omega_a)}$
      & $\begin{matrix} \\ \\ \end{matrix} 0 \begin{matrix} \\ \\ \end{matrix} $
\\ \hline
\end{tabular}}
\caption{Probabilities for the population of the possible helicity states of $\Lambda_c$, $\Sigma_c$ and $\Sigma_c^\star$.}
\label{tab4}
\end{table}
{\noindent Again, zero probabilities in Table~\ref{tab4} are the helicity states that are viable but vanishing with the initial assumption of $j_Q^{(3)}=-1/2$.} Additionally, the parameter $\omega_1$ can be estimated in a similar fashion as for $\omega_{3 \over 2}$ in Section~\ref{Mesons}, expanding and normalising Eq.~\eqref{diff_width} for the observed decay modes $\Sigma_c \to \Lambda_c \pi$ and $\Sigma_c^{\star} \to \Lambda_c \pi$:
\begin{equation}
{1 \over \Gamma}{d\Gamma(\Sigma_c \to \Lambda_c \pi) \over d\cos\theta} = {1 \over 2}\ ,
\end{equation} 
\begin{equation}
{1 \over \Gamma}{d\Gamma(\Sigma_c^{\star} \to \Lambda_c \pi) \over d\cos\theta} = {1 \over 4} 
\left[ 1 + 3 \cos^2 \theta - {9 \over 2} \omega_{1} \left( \cos^2 \theta - {1 \over 3} \right)\right],
\end{equation} 
that results in $\omega_{1} = {2/3}$. 

The above arguments can be trivially generalised for arbitrary heavy quark helicity:
\begin{table}[H]
\centering
\resizebox{\textwidth}{!}{%
\begin{tabular}{|c | | c c c c|}
\hline
$\hat{\rho}$ & $\rho_{0,0}$ & $\rho_{1,1}$ & $\rho_{2,2}$ & $\rho_{3,3}$ 
\\ \hline \hline
$\Lambda_c$ & ${1\over 2}(1-\rho_Q)$
      & ${1\over 2}(1+\rho_Q)$
      & $-$
      & $\begin{matrix}  \\ \\ \end{matrix} -\begin{matrix}  \\ \\ \end{matrix}$ 
\\ 
$\Sigma_c$ & ${1\over 2}(1-\rho_Q) + \omega_{1} \rho_Q$
       & ${1\over 2}(1+\rho_Q) - \omega_{1} \rho_Q$
       & $-$
       & $\begin{matrix} \\ \\ \end{matrix}-\begin{matrix}  \\ \\ \end{matrix}$ 
\\ 
$\Sigma_c^\star$ & ${3\over 8}\omega_{1}(1-\rho_Q)$
      & ${1\over 2}(1-\rho_Q) - {1\over 8}\omega_{1}(3-5\rho_Q)$
      & ${1\over 2}(1-\rho_Q) - {1\over 8}\omega_{1}(3+5\rho_Q)$
      & $\begin{matrix} \\ \\ \end{matrix} 
        {3\over 8}\omega_{1}(1+\rho_Q)
        \begin{matrix} \\ \\ \end{matrix} $
\\ \hline
\end{tabular}}
\caption{Possible polarisation states of charmed baryons, $\Lambda_c$, $\Sigma_c$ and $\Sigma_c^\star$.}
\label{tab5}
\end{table}
{\noindent Afterward, we are able to pass on the polarisation information of the heavy baryons in the transitional \texttt{SpinHadronizer}, through the \texttt{baryonSpin} sub-module. Commencing with checks on the baryon's parentage and its constituents, the module utilises spin information from the heavy quark to establish the spin characteristics of the baryon, especially relevant to excited heavy baryons:
\begin{itemize}

\item[(i)] $\Lambda_c^{+} ,\; \Lambda_c^{0} ,\; \Sigma_c^{++} ,\; \Xi_c^{0} ,\; \Xi_c^{+} ,\; \Xi_c^{'0} \; {\rm and \; C.C.}$

\item[(ii)] $\Sigma_c^{+} ,\; \Sigma_c^{0} ,\; \Sigma_c^{+} ,\; \Sigma_c^{0} ,\; \Xi_c^{+} ,\; \Xi_c^{0} ,\; \Omega_c^{0} ,\; \Omega_c^{0} \; {\rm and \; C.C.}$

\item[(iii)] $\Sigma_c^{++} ,\; \Xi_c^{++} ,\; \Xi_c^{+} ,\; \Omega_c^{+} ,\; \Xi_c^{+} ,\; \Xi_c^{++} ,\; \Omega_c^{++} ,\; \Omega_c^{+} ,\; \Xi_c^{+} \; {\rm and \; C.C.}$

\end{itemize}
The above items respectively correspond to rows 1 to 3 of Table~\ref{tab5}.}

\section{Decay of the Excited Heavy Mesons}
\label{sec:decay}

In the preceding section, we delineated a systematic methodology for transmitting the spin information of heavy hadron constituents, aiming to simulate the polarisation states of excited heavy mesons and heavy baryons within \textsf{Herwig7}, using the principles of HQET and spin-flavour symmetry. It is imperative to highlight that, given the limited available experimental data on these hadrons, our reliance on HQET to determine their decay modes, widths, and branching ratios is considerable. Notably, HQET does more than just earmark potential decay modes; it also provides a foundational basis for relationships among the associated couplings. This amplifies the importance of accurate modelling of these hadron decay modes. Historically, HQET has been utilised for the strong and radiative decays of heavy baryons, as evidenced by works in references\cite{Ivanov:1999bk,Ivanov:1998wj} and its incorporation into \textsf{Herwig++}~\cite{Bahr:2008pv,Bellm:2015jjp}. Yet, there remains a pressing need to model the decays of excited heavy mesons to optimally align predictions with the handful of extant experimental observations.

Considering the charmed mesons, we can identify the multiples based on our initial categorisation: (i) the ground state $D$ and $D^{\star}$ mesons, (ii) the $D_1$ and $D_2^{\star}$ doublets and (iii) the $D_0^{\star}$ and $D_1^\prime$, doublets. To determine the relevant matrix elements for the decay of these charmed mesons, we follow the methodology and notation detailed in Refs.~\cite{Falk:1992cx,Falk:1995th}. We prioritise the leading terms within the heavy quark limit for the interactions, while also considering the mixing between the $D_1$ and $D_1^\prime$ mesons, $\theta_q$. This yields the following expression for the decay matrix elements:
\begin{align}
\mathcal{M}(D^{\star}\to D\pi) &= -{2g \over f_\pi} \left( m_Dm_{D^{\star}} \right)^{1\over 2} p_0\cdot \epsilon_0 ,
\label{m1}
\\
\mathcal{M}(D_2^{\star}\to D\pi) &= -{2h \over f_\pi \Lambda} \left( m_{D_2}m_D^{\star} \right)^{1\over 2} \epsilon_0^{\mu\nu} p_{0,\mu} p_{0,\nu} ,
\label{m2}
\\
\mathcal{M}(D_2^{\star}\to D^{\star}\pi) &= -i{2h \over f_\pi \Lambda} \left( {m_{D^{\star}} \over m_{D_2}} \right)^{1\over 2}  \epsilon^{\alpha\beta\mu\nu} \epsilon^0_{\alpha\gamma} p_0^\gamma p_{0,\mu} p_{1\nu}\epsilon_{1\beta} ,
\label{m3}
\\
\mathcal{M}(D_1\to D^{\star}\pi) &=  {h \over f_\pi\Lambda}   \left( {2 \over 3} {m_{D_1}m_D} \right)^{1\over 2}\left[ \epsilon_0\cdot\epsilon_1\left(p_0^2-\left[{p_0\cdot p_1 \over m_0}\right]^2\right)-3\epsilon_0\cdot p_0\epsilon_1\cdot p_0\right] ,
\label{m4}
\\
\mathcal{M}(D_0^{\star}\to D\pi) &= {f^{\prime\prime} \over f_\pi}\left( {m_{D_0^{\star}}m_D} \right)^{1\over 2} \,p_0\cdot\left({p_1 \over m_{D^{\star}_0}}+{p_2 \over m_D}\right) ,
\label{m5}
\\
\mathcal{M}(D_1^\prime \to D^{\star}\pi) &=-{f^{\prime\prime} \over f_\pi}\left( {m_{D_1^\prime}m_D} \right)^{1\over 2}\left[-p_0\cdot\left({p_1 \over m_{D^{\star}_0}}+{p_2 \over m_D}\right)\epsilon_0\cdot\epsilon_1\right. 
  \nonumber\\
&  \left. +{1 \over m_{D'_1}}\epsilon_1\cdot p_1\epsilon_0\cdot p_0
+{1 \over m_{D}}\epsilon_0\cdot p_2\epsilon_1\cdot p_0 \right].
\label{m6}
\end{align}
Here, $p_i$ and $\epsilon_i$ are the momenta and the polarisation vectors of the hadrons, with $i=0$ pointing to the parent hadron and $i=1,2$ to the heavy and light child hadrons respectively. $m_H$ is the mass of hadron $H$ and $g$, $h$, $\Lambda$, $f_{\pi}$ and $f^{''}$ are decay parameters. We also introduce $\theta_q$ as the mixing angles between ($D_1$, $D'_1$) and ($D_{s1}$, $D'_{s1}$) mesons.

Having the matrix elements of the decays, we can calculate the partial widths using the Fermi golden rule for two-body decays,
\begin{equation}
\Gamma(H^{\star} \to H \pi) = {\Theta(H^{\star} \to H \pi) \over 2m_{H^{\star}}} 
\left| \mathcal{M}(H^{\star} \to H \pi) \right|^2,
\end{equation}
with $\Theta$ being the two-body phase-space factor. Furthermore, to ensure consistency between the theoretical calculations and our implementation in \textsf{Herwig~7}, it becomes necessary to retain some of the sub-leading terms when considering the heavy quark limit. Explicitly, we use the following prescription for our calculations:
\begin{equation}
\Gamma(H^{\star} \to H \pi) = {1 \over 8 \pi \; m_{H^{\star}}^2} 
\left| \mathcal{M}(H^{\star} \to H \pi) \right|^2 p_{\rm CM},
\end{equation}
where $p_{\rm CM}$ denotes the momentum of the parent hadron in the kinematical centre-of-mass frame of the two-body decay. The resulting partial widths are outlined below:
\begin{subequations}
  \begin{align}
    \Gamma (D^{\star}\to D\pi) & =  \frac{g^2}{6\pi f^2_\pi}  \frac{m_D}{m_{D^{\star}}} p_{\text{cm}}^3 \\
    \Gamma (D_2^{\star}\to D^{\star}\pi) & = \frac{h^2}{15\pi f^2_\pi\Lambda^2} \frac{m_{D^{\star}}}{m_{D^{\star}_2}} p_{\text{cm}}^5\\
    \Gamma (D_2^{\star}\to D\pi) &=  \frac{h^2}{10 \pi f^2_\pi\Lambda^2}  \frac{m_{D}}{m_{D^{\star}_2}}  p_{\text{cm}}^5\\
    \Gamma (D_1\to D^{\star}\pi) &=  \frac{h^2}{144 \pi f^2_\pi\Lambda^2}  \frac{\left[-2m^2_{D^1} (m^2_{D^{\star}}-5 m_\pi^2)+(m_\pi^2-m^2_{D^{\star}})^2+25 m^4_{D^1}\right]}{m_{D^{\star}}m^3_{D^1}}   p_{\text{cm}}^5  \\
    \Gamma (D_0^{\star}\to D\pi) &= \frac{ f^{\prime\prime2}}{32 f_\pi^2\pi}         \frac{\left(m_{D^{\star}_0}-m_{D}\right)^2     }{m_{D}m^3_{D^{\star}_0}} \left[ \left(m_{D^{\star}_0}+m_{D}\right)^2 -m_\pi^2\right]^2  p_{\text{cm}} \\
    \Gamma (D_1^\prime \to D^{\star}\pi) &= \frac{ f^{\prime\prime2}}{32 f_\pi^2\pi }  \frac{\left(m_{D^\prime_1}-m_{D^{\star}}\right)^2}{m_{D^{\star}}m^3_{D^\prime_1}}  \left[ \left(m_{D^\prime_0}+m_{D^{\star}}\right)^2 -m_\pi^2\right]^2p_{\text{cm}}
  \end{align}
\end{subequations}
Note that in \textsf{Herwig~7} uses both the amplitude and the decay width expressions simulation of the decay modes to form a precise phase space.  

To extract the numerical values of the decay parameters $g$, $h$, $\Lambda$, $f_{\pi}$ and $f^{''}$ in Eqs.~\eqref{m1} through~\eqref{m6}, we make use of the latest observed masses and decay widths of the charmed mesons from BaBar~\cite{BaBar:2010zpy,BaBar:2011vbs} and LHCb~\cite{LHCb:2013jjb} collaborations. The complete list of utilised masses and widths for the $(0^-,1^-)$, $(1^+,2^+)$ and $(0^+,1^+)$  multiplets are given in Table~\ref{tab6}. 
\begin{table}[H]
\centering
\resizebox{\textwidth}{!}{%
\begin{tabular}{|c||c||c c c||c|}
\hline
Multiplet & State & Meson & Mass [GeV] & Width & $\Delta m$ \\
\hline \hline
$(0^-,1^-)$ & $0^-$ & $D^+$      & $ 1.8697 \pm  0.0001$ & N/A & $   4.82 \pm    0.07$\\
            &       & $D^0$      & $ 1.8648 \pm  0.0001$ & N/A & $   0.00 \pm    0.00$\\
            &       & $D^+_s$    & $ 1.9683 \pm  0.0001$ & N/A & $ 103.51 \pm    0.09$\\
\cline{2-6}
            & $1^-$ & $D^{*+}$   & $ 2.0103 \pm  0.0001$ & $( 0.0834 \pm  0.0018)\times10^{-3}$ & $   3.41 \pm    0.07$\\
            &       & $D^{*0}$   & $ 2.0069 \pm  0.0001$ & $< 0.0021$ & $   0.00 \pm    0.00$\\
            &       & $D^{*+}_s$ & $ 2.1122 \pm  0.0004$ & $< 0.0019$  & $ 105.35 \pm    0.40$\\
\hline \hline
$(1^+,2^+)$ & $1^+$ & $D_1^+$      & $ 2.4232 \pm  0.0024$ & $ 0.0250 \pm  0.0060$ & $   2.40 \pm    2.45$\\
            &       & $D_1^0$      & $ 2.4208 \pm  0.0005$ & $ 0.0317 \pm  0.0025$ & $   0.00 \pm    0.00$\\
            &       & $D_{s1}^+$   & $ 2.5351 \pm  0.0001$ & $ 0.0009 \pm  0.0001$ & $ 114.31 \pm    0.50$\\
\cline{2-6}
            & $2^+$ & $D_2^{*+}$      & $ 2.4654 \pm  0.0013$ & $ 0.0467 \pm  0.0012$ & $   4.70 \pm    1.36$\\
            &       & $D_2^{*0}$      & $ 2.4607 \pm  0.0004$ & $ 0.0475 \pm  0.0011$ & $   0.00 \pm    0.00$\\
            &       & $D_{s2}^{*+}$   & $ 2.5691 \pm  0.0008$ & $ 0.0169 \pm  0.0007$ & $ 108.40 \pm    0.89$\\
\hline \hline
$(0^+,1^+)$ & $0^+$ & $D_0^+$      & $ 2.3490 \pm  0.0070$ & $ 0.2210 \pm  0.0180$ & $  49.00 \pm   20.25$\\
            &       & $D_0^0$      & $ 2.3000 \pm  0.0190$ & $ 0.2740 \pm  0.0400$ & $   0.00 \pm    0.00$\\
            &       & $D_{s0}^+$   & $ 2.3178 \pm  0.0005$ & $< 0.0038$ & $  17.80 \pm   19.01$\\
\cline{2-6}
            & $1^+$ & $D_1^{\prime+}$  &\multicolumn{2}{c|}{Not observed}     & \\
            &       & $D_1^{\prime0}$      & $ 2.4270 \pm  0.0400$ & $   0.38^{+   0.13}_{-   0.11}$ & $   0.00 \pm    0.00$\\
            &       & $D_{s1}^{\prime+}$   & $ 2.4595 \pm  0.0006$ & $< 0.0035$ & $  32.50 \pm   40.00$\\
\hline
\end{tabular}}
\caption{Observed masses and widths of the charm mesons, from BaBar~\cite{BaBar:2010zpy,BaBar:2011vbs} and LHCb~\cite{LHCb:2013jjb} collaborations.}
\label{tab6}
\end{table}

\begin{figure}[H]
\includegraphics[width=0.5\textwidth]{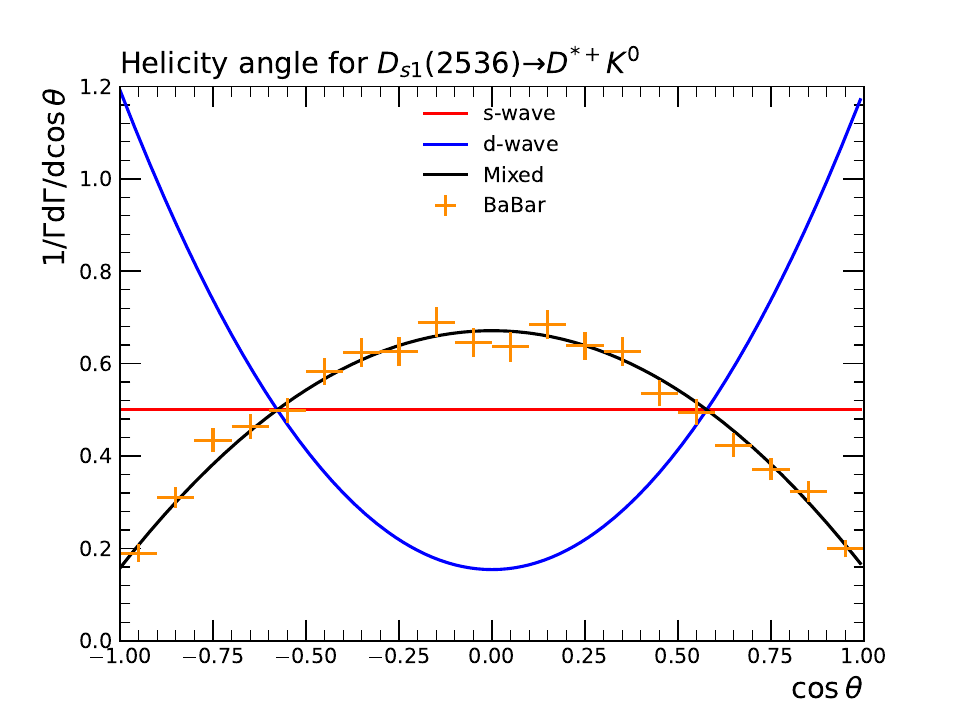}\hfill
\includegraphics[width=0.5\textwidth]{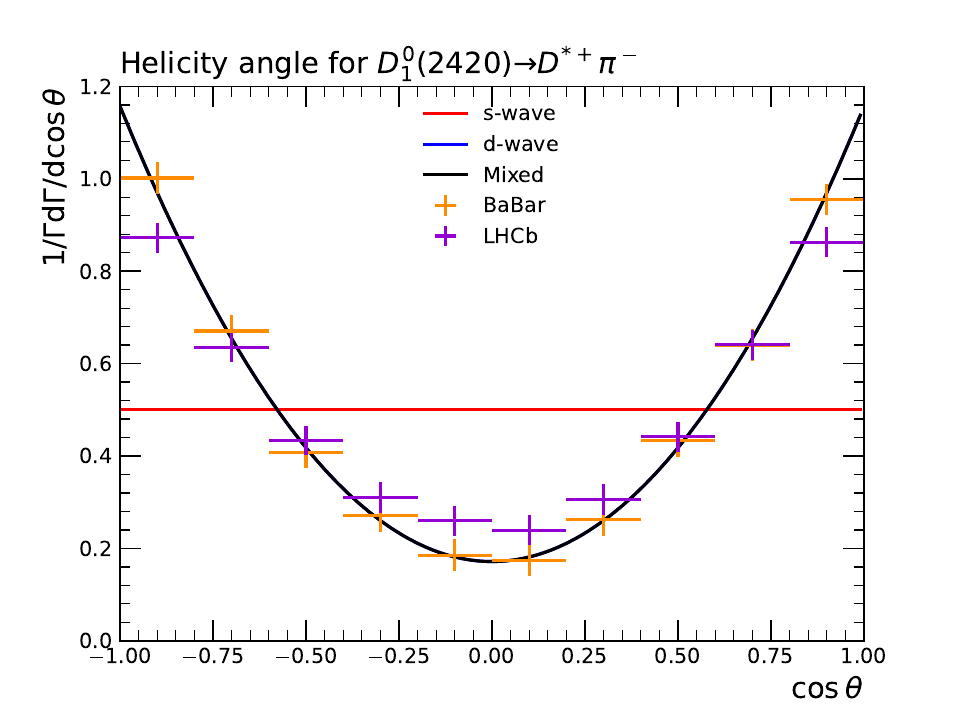}\\
\caption{Helicity Angle distributions for the decays: (left panel) $D_{s1}(2536)\to D^{*+}K^0$ and (right panel) $D_{1}^0(2420)\to D^{*+}\pi^-$.}
\label{fig1}
\end{figure}

Figure~\ref{fig1} shows the helicity angle distributions for the $D_{s1}(2536)\to D^{*+}K^0$ and $D_{1}^0(2420)\to D^{*+}\pi^-$ decays modes. $s$- and $d$-wave contributions are shown independently and the mixed sums are fitted against the corresponding experimental data. The fitting is performed using \textsf{Minuit} python module~\cite{James:1975dr} to find the best values for the $g$, $h$, $\Lambda$, $f_{\pi}$, $f^{''}$ and $\theta_q \;(q=u,d,s)$ parameters. These best-fit values are recorded in Table~\ref{tab7}. 
\begin{table}[H]
\centering
\resizebox{0.35\textwidth}{!}{%
\begin{tabular}{|c||c|}
\hline 
Parameter      & Fitted Value              \\ \hline \hline
$f''$          & $ -0.465 \pm 0.017$       \\ \hline
$f_\pi$        & $  0.130 \pm 0.001$ [GeV] \\ \hline
$h$            & $  0.824 \pm 0.007$       \\ \hline
$\Lambda$      & $  1.000 \pm 0.000$ [GeV] \\ \hline
$g$            & $  0.565 \pm 0.006$       \\ \hline
$\theta_{u,d}$ & $  0.000 \pm 0.100$       \\ \hline
$\theta_s$     & $ -0.047 \pm 0.002$       \\ \hline
\end{tabular}}
\caption{Fitted values of the decay parameters $g$, $h$, $\Lambda$, $f_{\pi}$, $f^{''}$ and $\theta_q \;(q=u,d,s)$, using the charm mesons masses and decay widths from Refs.~\cite{BaBar:2010zpy,BaBar:2011vbs,LHCb:2013jjb}.}
\label{tab7}
\end{table}

Besides the strong isospin-conserving decays, certain excited mesons also undergo electromagnetic decays, as well as strong-isospin-violating ones. These latter decays are particularly prominent in cases where the strong isospin-conserving decays are either kinematically inhibited or strongly suppressed, either because of the lack of sufficient rest energy, threshold effects, angular momentum conservation, particular selection rules in the given event handler, hierarchy of coupling constants or presence of other dominant channels. These are predominantly observed in the following cases:
\begin{itemize}
\item[(i)] $D^{\star}$ mesons: Here, the strong isospin-conserving decays are kinematically limited, making radiative modes crucial.

\item[(ii)] $B^{\star}$ mesons: In this case, the strong isospin-conserving decays cannot occur due to kinematic constraints. Thus, only the radiative mode emerges as a possibility.

\item[(iii)] $D^{+}_s$, $D^{+}_{s0}$, and $D^{+}_{s1}(2460)$ mesons: For these, both radiative and isospin violating decay modes hold significance since the strong isospin conserving $DK$ modes are kinematically proscribed.

\item[(iv)] $B_s^{*0}$ meson: Only the radiative mode is feasible from a kinematic perspective.
\end{itemize}
Again, our focus will be on the $D^{\star}$ system, as it represents the most intricate scenario, exhibiting a plethora of excited mesons below the strong decay threshold. Nevertheless, our arguments can be effortlessly extended to other cases. 

The amplitude for the radiative decay of the $D^{\star}$ mesons, as detailed in~\cite{Cho:1992nt}, is given by:
\begin{equation}
\mathcal{M}(D^{\star}\to D\gamma) = A \left[ 64\pi \frac{m_D}{m_{D^{\star}}} \right]^{1 \over 2} \epsilon^{\alpha\beta\mu\nu}\epsilon_{\gamma\alpha}p_{\gamma\beta}\epsilon_{D^{\star}\mu}p_{D^{\star}\nu}.
\end{equation}
Here, the coupling, $A$, is expressed as:
\begin{equation}
 A= \frac{e_Q}{4m_Q}\alpha(m_Q)^{\frac{1}{2}} +\frac{c_H}{\Lambda}e_q\alpha(\Lambda)^{\frac12},
\end{equation}
with $e_Q$ and $e_q$ being the electric charges of the heavy and light quarks respectively $m_Q$ being the mass of the heavy quark. $\alpha$ is the electromagnetic coupling constant while $c_H = -1.058$ is electromagnetic coefficient for heavy meson decays. The scale $\Lambda$ is the same as the case of strong decays. Consequently, the partial width becomes~\cite{Bardeen:2003kt,Cho:1994zu}:
\begin{equation}
\Gamma(D^{\star}\to D\gamma) = \frac{2 m_D|A|^2 }{3 m_{D^{\star}}} \left(\frac{m_{D^{\star}}^2-m_D^2}{m_{D^{\star}}}\right)^3.
\end{equation}

To implement the strong and radiative decays of excited heavy mesons in \textsf{Herwig~7}, we introduced two specialised classes: \texttt{HQETStrongDecayer} and \texttt{HQETRadiativeDecayer}. The decay parameters highlighted in this section are defined as user-adjustable variables. This design provides flexibility, facilitating potential tuning and refinement based on future insights or requirements. In the subsequent section, we will evaluate the robustness of our implementation by juxtaposing it against the available experimental data.

Here we note that the HQET treatment of heavy baryon decays is not discussed here since this has been implemented in \textsf{Herwig~7}, continuing from its initial implementation in \textsf{Herwig++}~\cite{Bahr:2008pv,Bellm:2015jjp}. This implementation includes the decays of heavy baryons such as $\Lambda_b$, $\Xi_b$, and $\Omega_b$, following the principles outlined in HQET. These decays are handled using the same theoretical framework applied to heavy mesons, ensuring consistency across the different types of heavy hadrons. The decay modes, branching ratios, and angular distributions for these baryons are calculated based on HQET, taking into account the relevant spin-flavour symmetries and mixing effects. This comprehensive implementation allows for accurate simulation of heavy baryon decays in \textsf{Herwig~7}, providing a valuable tool for studying these processes in high-energy physics experiments.

\section{Examination of the Observables}
\label{sec:hadronresults}

The integration of the \texttt{SpinHadronizer} into \textsf{Herwig~7} enabled more accurate polarisation details for excited heavy mesons and baryons, based on their constituent heavy quarks' spin. Additionally, the \texttt{HQETStrongDecayer} and \texttt{HQETRadiativeDecayer} classes were incorporated, further enhancing \textsf{Herwig}'s ability to model both strong and radiative decays, thereby providing a comprehensive framework for heavy hadron decay dynamics. To evaluate these improvements, we scrutinised existing experimental data, particularly those that are sensitive to hadron polarisation states. Limited but valuable insights were gained from measurements like the angular decomposition of the $D_{s1}^+ \to D^{\star +} K_{s0}$ decay mode, as observed by BaBar~\cite{BaBar:2011vbs} and BELLE~\cite{Belle:2007kff} and from $\Lambda_b$ baryon polarisation in hadronic $Z^0$ decays by ALEPH~\cite{ALEPH:1995aqx} and OPAL~\cite{OPAL:1998wmk}.

\begin{figure}[t]
\centering
\includegraphics[width=.49\textwidth]{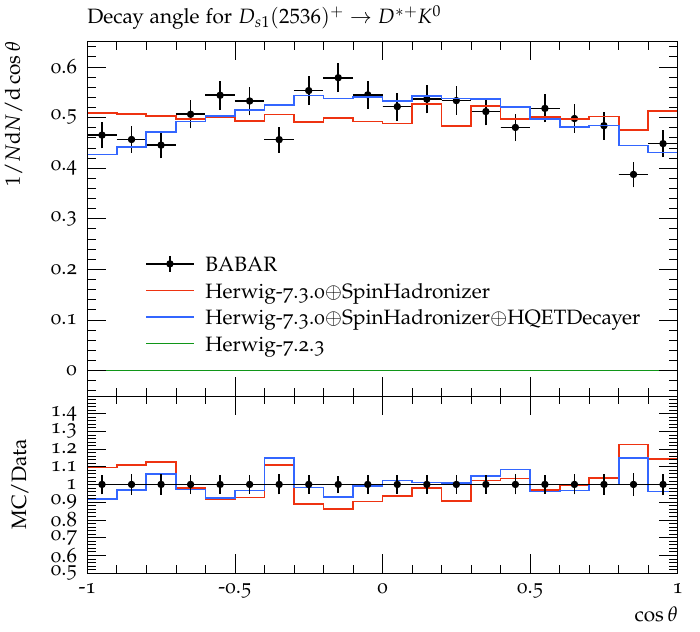}
\includegraphics[width=.49\textwidth]{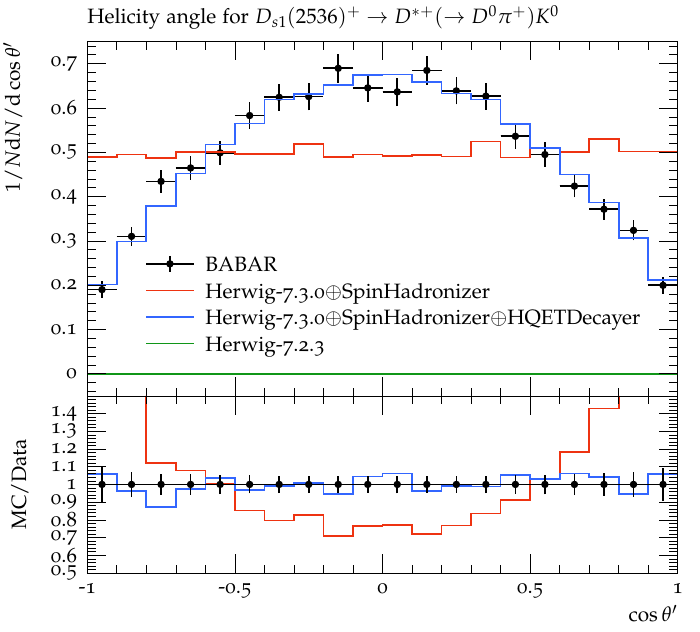}
\caption{Efficiency-corrected decay rates of the $D_{s1}^+$ meson: (left panel) displays the $D_{s1} \to D^{\star +} K^0$ decay mode rates as a function of lab-frame angle $\theta$. (right panel) shows the same but in the $D_{s1}^+$ center-of-mass frame, denoted by $\theta'$. Data sourced from Ref.~\cite{BaBar:2011vbs}. Red histograms represent \textsf{Herwig7.3.0} predictions with only the \texttt{SpinHadronizer} active, while blue histograms indicate the combined effect of \texttt{SpinHadronizer} and \texttt{HQETDecayer} classes. Green histograms correspond to \textsf{Herwig~7.2.3} predictions, which lack HQET enhancements.}
\label{fig2}
\end{figure}

\begin{figure}[t]
\centering
\includegraphics[width=.49\textwidth]{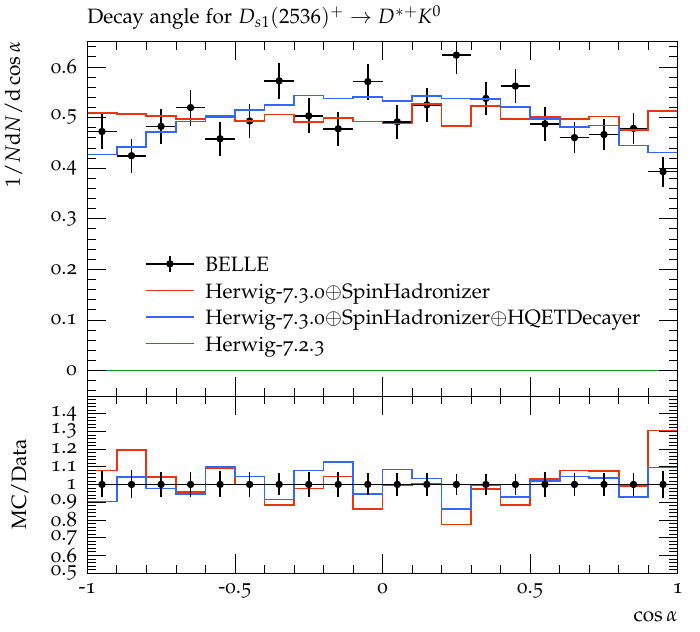}
\includegraphics[width=.49\textwidth]{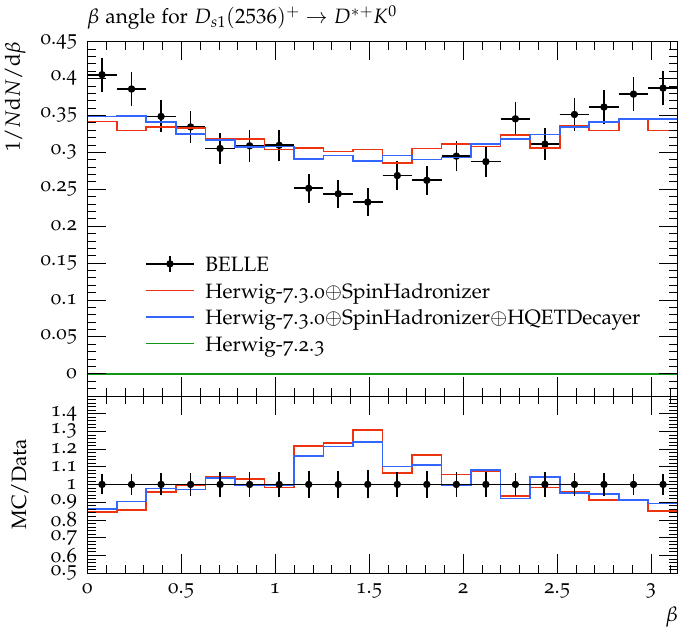}
\includegraphics[width=.49\textwidth]{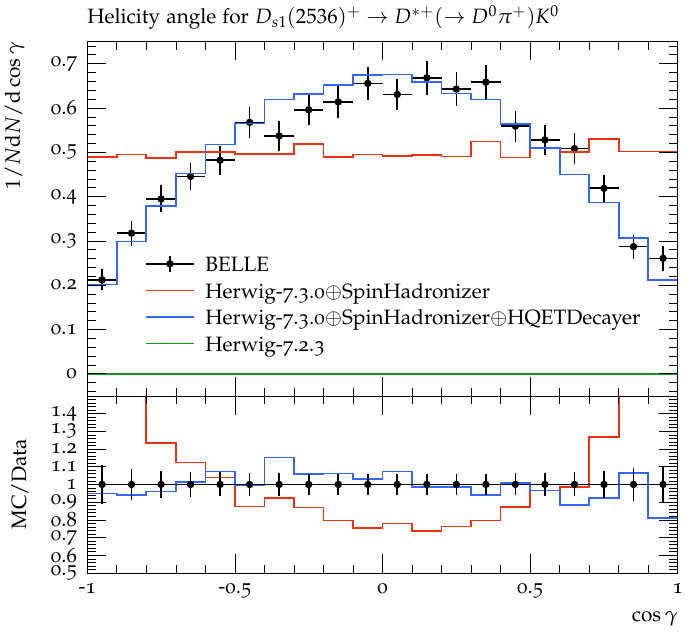}
\caption{Further analysis of efficiency-corrected decay rates for $D_{s1}^+$: (left panel) rates of $D_{s1} \to D^{\star +} K^0$ as a function of lab-frame angle $\alpha$. (right panel) the same rates with respect to angle $\beta$. (bottom panel) focuses on the angle $\gamma$ between $\pi^+$ and $K^0$ in the $D^{\star +}$ rest frame. The data is sourced from~\cite{Belle:2007kff} and the histograms are annotated as in Figure~\ref{fig2}. }
\label{fig3}
\end{figure}

In Figures~\ref{fig2}~and~\ref{fig3}, we investigate the efficacy of using HQET and spin-flavor symmetry to predict $e^-e^+$ data concerning polarisation-sensitive measurements of the $D_{s1}$ meson decays. The figures display normalised, efficiency-corrected rates for the $D_{s1} \to D^{\star +} K^0$ decay mode as functions of different decay angles. We evaluate the contributions of the \texttt{SpinHadronizer} and \texttt{HQETDecayer} classes in \textsf{Herwig-7.3.0}, activating them individually to study their distinct effects on the predictions. For comparison, we also include predictions from the previous version, \textsf{Herwig-7.2.3}, which lacks both spin-flavor symmetry and HQET enhancements. As anticipated, \textsf{Herwig-7.2.3} fails to accurately predict either the existence or the behaviour of angle-dependent decay rates. On the other hand, \textsf{Herwig-7.3.0} correctly predicts the mean values of the $s$-wave contributions when only the \texttt{SpinHadronizer} class is activated, but does not capture the full angle-dependent behaviour. Importantly, with both the \texttt{SpinHadronizer} and \texttt{HQETDecayer} classes enabled, \textsf{Herwig-7.3.0} achieves an excellent agreement with the observed data.

\begin{figure}[t]
\centering
\includegraphics[width=.49\textwidth]{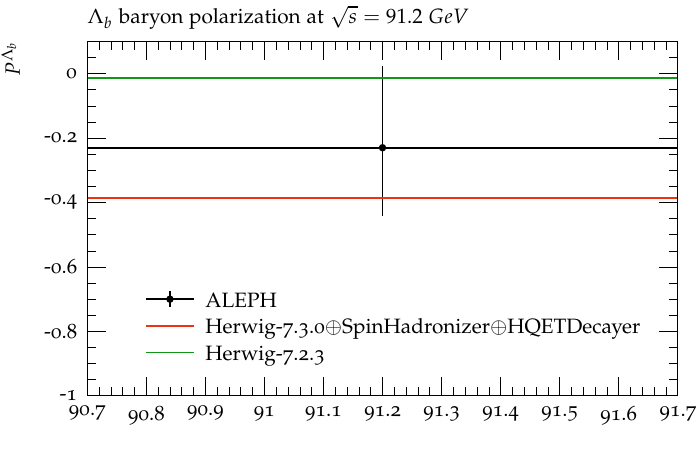}
\includegraphics[width=.49\textwidth]{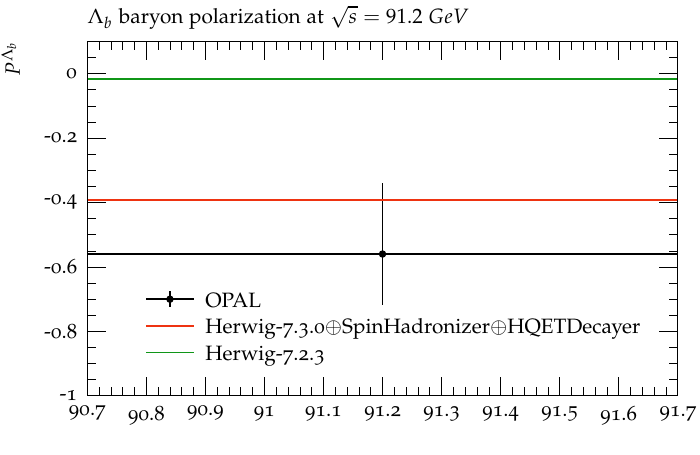}
\caption{Average polarisation of $\Lambda_b$ baryons in hadronic $Z^0$ decays at LEP by ALEPH~\cite{ALEPH:1995aqx} (left panel) and OPAL~\cite{OPAL:1998wmk} (right panel) for $\sqrt{s}=91.2$ GeV. Histograms are annotated as in Figure~\ref{fig2}.}
\label{fig_baryons}
\end{figure}

For heavy baryons, we validate our implementation of HQET and spin-flavor symmetry by comparing with measurements of the average polarisation of $\Lambda_b$ baryons in hadronic $Z^0$ decays at LEP, as reported by the ALEPH~\cite{ALEPH:1995aqx} and OPAL~\cite{OPAL:1998wmk} collaborations, shown in Figure~\ref{fig_baryons}. Notably, although the predictions from \textsf{Herwig-7.2.3} are small yet distinct from zero, the combined use of \texttt{SpinHadronizer} and \texttt{HQETDecayer} in \textsf{Herwig-7.3.0} leads to predictions nearly an order of magnitude larger, while consistent with the experimental uncertainty.

After incorporating modifications to the hadronization and decays of heavy excited mesons and heavy baryons in \textsf{Herwig 7.3.0}, it is advisable to revisit the production rates of these hadrons in the simulation. This ensures that the framework remains consistent, particularly if the new modifications could significantly influence kinematic distributions or other dependent observables. With this prospect in mind, we initially undertake a general tuning effort, trying to optimise the relevant existing and newly defined parameters to be in synergy with a large pool of experimental data. In the next section, we will describe our methodology for this general tune and its outcome, before re-evaluating the production rates of the heavy and excited hadrons for a rigorous validation process.

\section{General Tune for Herwig-7.3.0}
\label{sec:tune}

For the first general tune of \textsf{Herwig 7.3.0} since the 7.2.0 release, we utilised $e^-e^+$ data from LEP, PETRA, SLAC, SLC, and TRISTAN, spanning over 9,200 separate data bins. The data were weighted around both light and heavy hadron production rates and multiplicities, with a focus on more dominant processes. Initial attempts to use \textsf{Professor~II} yielded inconsistencies due to the complexity of the tuning and the high number of parameters. We then resorted to a multi-layered, brute-force approach using the \texttt{prof2-chisq} module from \textsf{Professor~II}, minimising the $\chi^2$ as an indicator of the best tune. This approach was conducted over five layers (each with 5000 randomly distributed samples), where each layer aimed to refine the parameters further based on the $\chi^2$ improvements observed in the previous layer. After each layer, the best parameter sample was chosen, and the parameter space was reduced by 50\% around that sample. The tuning stopped when the sixth layer did not produce better (or worse) $\chi^2$ values compared to the fifth.

In our tuning procedure, we use a heuristic $\chi^{2}$ function defined as:
\begin{equation}
\chi^{2}(p) = \sum_{\mathcal{O}} w_{\mathcal{O}} \sum_{b \in \mathcal{O}} \frac{\left(f_b(p) - \mathcal{R}_b\right)^2}{\Delta_b^2},
\end{equation}
where $p$ is the set of parameters being tuned, $\mathcal{O}$ are the observables each with weight $w_{\mathcal{O}}$, $b$ are the different bins in each observable distribution with associated experimental measurement $\mathcal{R}_b$, error $\Delta_b$, and Monte-Carlo prediction $f_b(p)$. To ensure that key observables significantly influence the result of the fit, we use different weights for different types of data. Specifically, we use $w_{\mathcal{O}}=1$ in most cases, $w_{\mathcal{O}}=10$ for particle multiplicities, $w_{\mathcal{O}}=50$ for total charged particle multiplicities, and $w_{\mathcal{O}}=100$ for hadron production rates. This weighting ensures that particle multiplicities and hadron production rates have a substantial impact on the fit due to the higher quantity of event shape and spectrum data used in the tuning. Additionally, we apply $w_{\mathcal{O}}=10$ for data on gluon jets to prevent the fit from being dominated by the large quantity of data sensitive to quark jets. To avoid the fit being dominated by a few observables with very small experimental errors, which could lead to overfitting, we use an effective error defined as:
\begin{equation}
\Delta^{\rm eff}_b = \max(0.05 \times \mathcal{R}_b, \Delta_b),
\end{equation}
instead of the true experimental error $\Delta_b$ in the fit. This approach ensures that the fit remains robust and not overly sensitive to individual data points with extremely small uncertainties.

In this context, we analysed a total of 12 parameters, comprising 10 for cluster hadronization and 2 for the parton shower---specifically \texttt{AlphaIn} and \texttt{pTmin}. Below is a detailed breakdown of these parameters:
\begin{itemize}
    \item \texttt{ClMaxLight}: Maximum allowable cluster mass for light quarks.
    \item \texttt{ClPowLight}: Power exponent for the mass of clusters with light quarks.
    \item \texttt{PSplitLight}: Parameter affecting the mass splitting for clusters with light quarks.
    \item \texttt{PwtSquark}: Probability for a $s\bar{s}$ quark pair to be spawned during cluster splittings.
    \item \texttt{PwtDIquark}: Probability for quarks forming a di-quark.
    \item \texttt{SngWt}: Weighting factor for singlet baryons in hadronization.
    \item \texttt{DecWt}: Weighting factor for decouplet baryons in hadronization.
    \item \texttt{ProbabilityPowerFactor}: Exponential factor in the \texttt{ClusterFissioner} probability function.
    \item \texttt{ProbabilityShift}: Offset in the \texttt{ClusterFissioner} probability function.
    \item \texttt{KinematicThresholdShift}: Adjustment to the Kinematic threshold in \texttt{ClusterFissioner}.
    \item \texttt{AlphaIn}: Initial value for the strong coupling constant at $M_{Z^0}=91.1876$ GeV\footnote{The reader should note that after the \textsf{Herwig-7.3.0} release, the \texttt{AlphaIn} has been split into two separate parameters, \texttt{AlphaQCD:AlphaIn} and \texttt{AlphaQCDFSR:AlphaIn} that control the strong coupling constants for initial- and final-state radiations respectively. Henceforth, the tuned value we suggest in Table~\ref{tuned_parameters} points to \texttt{AlphaQCDFSR:AlphaIn}, as it is more relevant to our $e^+e^-$ tuning strategy. Meanwhile, the tune proposed in \cite{Bewick:2019rbu,Bewick:2021nhc} has been adopted for the intial-state radiation coupling constant \texttt{AlphaQCD:AlphaIn}=$0.1185$.}.
    \item \texttt{pTmin}: Minimum transverse momentum in parton shower.
\end{itemize}
Notably, the parameters \texttt{ProbabilityPowerFactor}, \texttt{ProbabilityShift}, and \texttt{KinematicThresholdShift} are introduced only in \textsf{Herwig 7.3.0}. The first two parameters belong to the newly introduced \texttt{ClusterFissioner} Probability Function, $P_{\rm cluster}$, which serves as a decision-making utility that evaluates whether a given cluster meets a dynamically calculated kinematic threshold:
\begin{equation}
P_{\rm cluster} = \frac{1}{1 + \left| \frac{M - \delta}{M_{\rm th}} \right|^r}, \qquad P_{\rm cluster} > P_{\rm random}(0,1),
\end{equation}
Here, $M$ denotes the effective mass of the cluster while $M_{\rm th}$ is the mass threshold for the cluster, defined as the sum of the masses of the constituent quarks plus the mass of the spawned di-quark. $\delta$ and $r$ are the \texttt{ProbabilityShift} and \texttt{ProbabilityPowerFactor} parameters, respectively. $P_{\rm cluster}$ is then compared against a randomly generated number between 0 and 1, denoted as $P_{\rm random}(0,1)$, to either accept or reject the cluster splitting. This new formulation provides a smooth Gaussian distribution for dynamic threshold cuts based on the cluster mass and its kinematic properties. Furthermore, the \texttt{Static} kinematic threshold for 2-body cluster decay in \textsf{Herwig-7.2.3} was:
\begin{equation}
M > M_1 + M_2, \quad M_1 >  m + m_1, \quad M_2 > m + m_2,
\end{equation}
This has been updated to a new, \texttt{Dynamic} threshold in \textsf{Herwig-7.3.0}:
\begin{equation}
M^2 > M_1^2 + M_2^2 , \quad M_1^2 > m^2 + m_1^2 + \delta_{\rm th}, \quad M_2^2 > m^2 + m_2^2 + \delta_{\rm th}.
\end{equation}
In this new formulation, $M_1$ and $M_2$ are the masses of the child clusters, $m_1$ and $m_2$ are the masses of the constituent quarks of the parent cluster, and $m$ represents the mass of the spawned quark-antiquark pair. $\delta_{\rm th}$ is the \texttt{KinematicThresholdShift} parameter. Note that both \texttt{Static} and dynamic threshold options are available in \textsf{Herwig-7.3.0}, with the default being the \texttt{Dynamic} choice:
\mybox{
	\texttt{cd /Herwig/Hadronization}
	\\
	\texttt{set  ClusterFissioner:KinematicThreshold <Static/Dynamic>}
	  }
	  
Following the above description and methodology, we tuned the hadronization and parton shower parameters in \textsf{Herwig 7.3.0}. The results are given in Table~\ref{tuned_parameters}. 
\begin{table}[H]
\centering
\resizebox{0.58\textwidth}{!}{%
\begin{tabular}{|c||c c|}
 \hline
Tuned Parameter & \textsf{Herwig-7.3.0} & \textsf{Herwig-7.2.0} \\
\hline\hline
\texttt{ClMaxLight} [GeV]& 3.529 & 3.649 \\
\texttt{ClPowLight} & 1.849 & 2.780 \\
\texttt{PSplitLight} & 0.914 & 0.899 \\
\texttt{PwtSquark} & 0.374 & 0.292 \\
\texttt{PwtDIquark} & 0.331 & 0.298 \\
\texttt{SngWt} & 0.891 & 0.740 \\
\texttt{DecWt} & 0.416 & 0.620 \\
\texttt{ProbabilityPowerFactor} & 6.486 & $-$ \\
\texttt{ProbabilityShift} & -0.879 & $-$ \\
\texttt{KinematicThresholdShift} [GeV]& 0.088 & $-$ \\
\texttt{AlphaIn} & 0.102 & 0.126 \\
\texttt{pTmin} [GeV]& 0.655 & 0.958 \\
\hline
\end{tabular}}
\caption{The values of tuned parameters in \textsf{Herwig 7.3.0} compared to \textsf{Herwig 7.2.0}.}
\label{tuned_parameters}
\end{table}

Despite its computational intensity, our brute-force approach can be deemed a reasonable success, achieving a substantial improvement in the overall $\chi^2$ values. Specifically, the tuned version of \textsf{Herwig 7.3.0} shows a $\sim$50.75\% improvement in $\chi^2$ compared to its untuned counterpart and a $\sim$12.76\% improvement when compared to \textsf{Herwig-7.2.3}. In the following section, we will look into the effects of this new tune on the production rates of heavy mesons and baryons.

\section{Heavy Hadron Production Rates}
\label{sec:rate}

In the realm of heavy hadron experimental physics, most of the available data on the production of excited states primarily stems from experiments focusing on the production of $B$ mesons from the $\Upsilon(4S)$ resonance, captured through data in the continuum region, i.e. at an energy range just above the production threshold for $B$ mesons but below the $\Upsilon(4S)$ peak, where $B$ mesons can be produced directly through $e^+e^-$ annihilation without necessarily forming a $\Upsilon$ state\footnote{This is in contrast to the ``resonance region", which is the energy range near the $\Upsilon(4S)$ peak, where the $e^+e^-$ collision is likely to form an $\Upsilon(4S)$  state before subsequently decaying into $B$ mesons.}. To thoroughly evaluate these phenomena, we undertook a substantial effort to collect and compile all accessible experimental data~\cite{Albrecht:1988dj,Albrecht:1989pa,Albrecht:1989yi,Albrecht:1991ss,Alexander:1993nq,Albrecht:1992zh,Albrecht:1995qx,Avery:1994yc,Bergfeld:1994af,Artuso:2004pj,Seuster:2005tr,Aubert:2002ue,Briere:2000np,Aubert:2006bk,Albrecht:1990zk,Albrecht:1997qa,Albrecht:1993pt,Albrecht:1988gc,Alexander:1999ud,Aubert:2005cu,Aubert:2006cp,Aubert:2006rv,Aubert:2006je,Aubert:2007bt,Avery:1990bc,Avery:1995ps,Bowcock:1989qh,Brandenburg:1996jc,Edwards:1994ar,Gibbons:1996yv,Jessop:1998wt,Mizuk:2004yu,Niiyama:2017wpp,Aubert:2001pd,Abe:2001za} into a single \textsf{Rivet} analysis, see Appendix~\ref{sec:AppDecay}. This comprehensive dataset served as a benchmark to scrutinise the production rates of all heavy mesons and baryons in \textsf{Herwig-7.3.0}. Particular attention was devoted to iso-spin multiplets; we meticulously calibrated their production rates to be nearly identical, thereby ensuring a robust and accurate hadronization and decay framework within the \textsf{Herwig-7.3.0} simulation.

\begin{figure}
\centering
\includegraphics[width=1\textwidth]{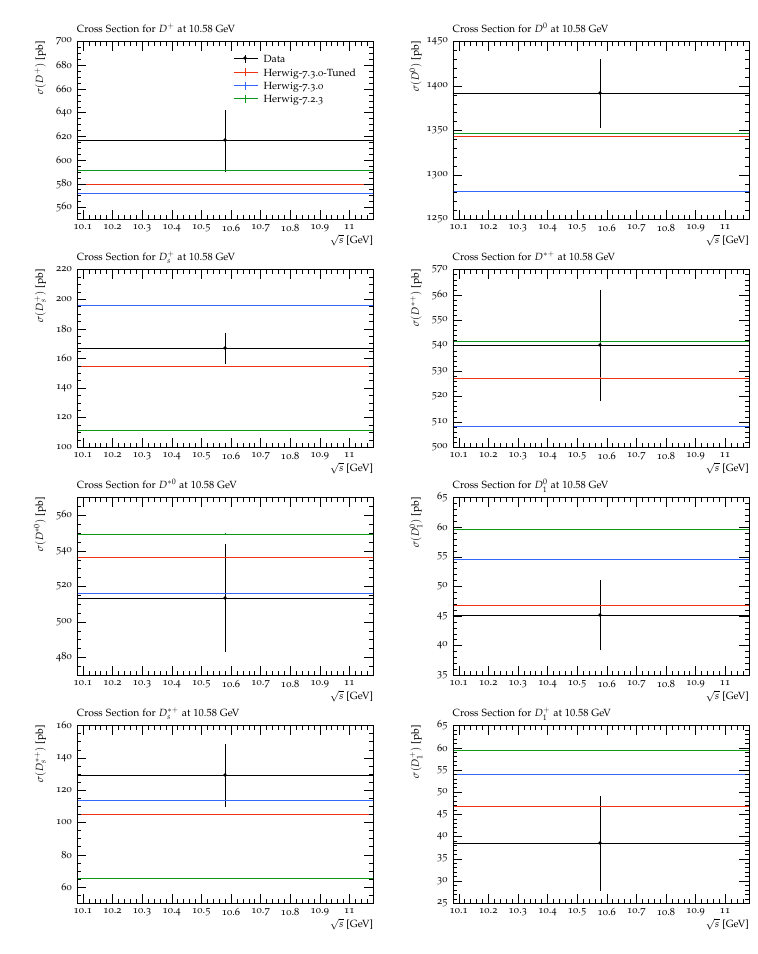}
\caption{Comparison of heavy meson production rates, collated from various experimental data sets~\cite{Albrecht:1988dj,Albrecht:1989pa,Albrecht:1989yi,Albrecht:1991ss,Alexander:1993nq,Albrecht:1992zh,Albrecht:1995qx,Avery:1994yc,Bergfeld:1994af,Artuso:2004pj,Seuster:2005tr,Aubert:2002ue,Briere:2000np,Aubert:2006bk}. The observed rates are compared against the predictions from both the tuned (represented by red lines) and untuned (depicted by blue lines) versions of \textsf{Herwig-7.3.0}. For reference, the results from \textsf{Herwig-7.2.3} are indicated by green lines.}
\label{fig4}
\end{figure}

\begin{figure}
\centering
\includegraphics[width=1\textwidth]{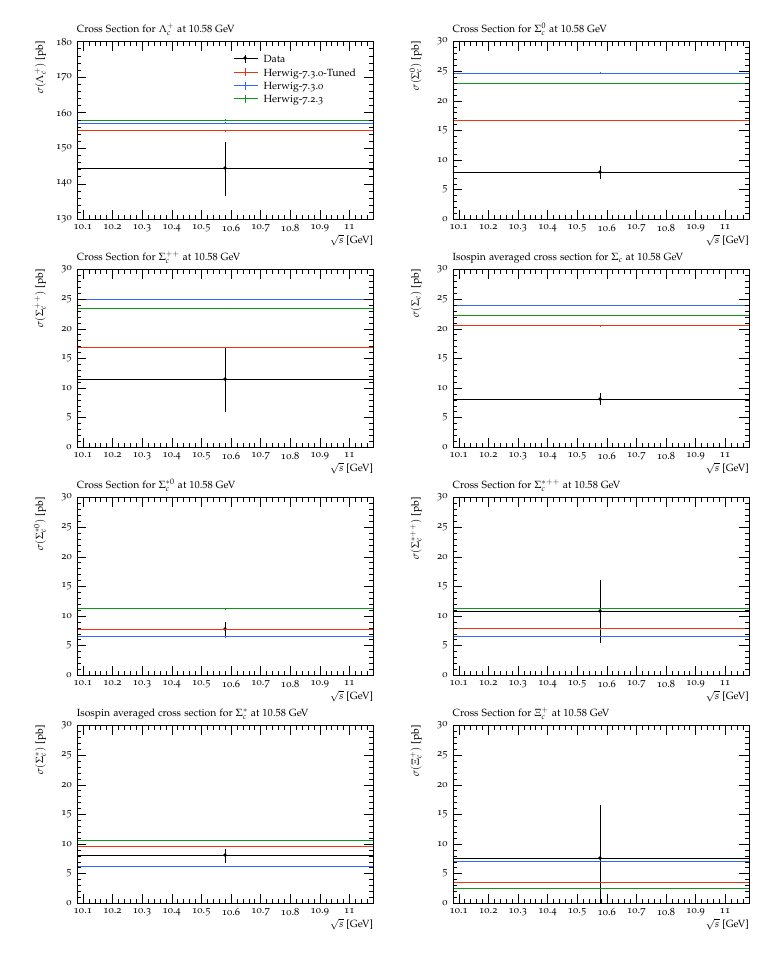}
\caption{Comparison of heavy baryon production rates, collated from various experimental data sets~\cite{Albrecht:1990zk,Albrecht:1997qa,Albrecht:1993pt,Albrecht:1988gc,Alexander:1999ud,Aubert:2005cu,Aubert:2006cp,Aubert:2006rv,Aubert:2006je,Aubert:2007bt,Avery:1990bc,Avery:1995ps,Bowcock:1989qh,Brandenburg:1996jc,Edwards:1994ar,Gibbons:1996yv,Jessop:1998wt,Mizuk:2004yu,Niiyama:2017wpp}. Histograms are annotated as in Figure~\ref{fig4}.}
\label{fig5}
\end{figure}

Figures~\ref{fig4} and~\ref{fig5} illustrate the impact of incorporating our spin-flavour symmetry approach into \textsf{Herwig-7.3.0} and HQET hadronic decay prescriptions. These figures compare the production rates of heavy mesons and baryons, as predicted by both the tuned and untuned versions of \textsf{Herwig-7.3.0}. For a comprehensive view, we also include corresponding results from \textsf{Herwig-7.2.3}. A notable enhancement in predictive accuracy is evident in the tuned \textsf{Herwig-7.3.0} simulations. 

The improvements introduced in the previous sections, combined with the new general tune of \textsf{Herwig-7.3.0} have made tangible improvements in the predictive modelling of heavy hadron production. This updated framework is poised to serve as an invaluable asset for both theoretical and experimental research in the high-energy physics community, especially for those focused on the complex dynamics of heavy hadron formation.

\section{Conclusion}
\label{sec:conc}

In this study, we have made substantial advancements in the predictive modelling of heavy hadron production through targeted modifications to the \textsf{Herwig~7} event generator. Specifically, we have overhauled its native cluster hadronization model by integrating HQET and spin-flavour symmetry. This was realised by adding three key modules: the \texttt{SpinHadronizer} class, which delivers polarisation information for excited heavy mesons and heavy baryons based on their constituent heavy quarks' helicity states; and the \texttt{HQETStrongDecayer} and \texttt{HQETRadiativeDecayer} classes, which improve the handling of heavy meson decays by enhancing their two-body decay matrix elements. This effectively complements the pre-existing heavy baryon decayers. Additionally, we formulated a new dynamic kinematic regime and included a decision-making probability function for smoother cluster splitting distributions, which improves the overall handling power of \textsf{Herwig~7} in the domain of heavy hadron production.

In the next phase of our efforts, we tuned \textsf{Herwig-7.3.0} using a brute-force, multilayered approach aimed at minimising the $\chi^2$ statistical parameter. To facilitate this, we assembled an extensive collection of experimental data tailored to our specific objectives. This data-driven methodology allowed us to fine-tune 12 parameters concurrently, including two parton shower-related parameters essential for the integration of the newly introduced electroweak parton shower. Consequently, we have achieved a notable elevation in \textsf{Herwig~7}'s capacity to make accurate predictions, particularly for the complex case of heavy hadron production.

The inclusion of HQET and spin-flavour symmetry, underpinned by robust theoretical foundations, serves as a significant milestone in improving the predictive capabilities of the cluster hadronization model. These methodological enhancements are poised to become invaluable assets for both theoretical and experimental research in high-energy physics, especially in areas requiring a nuanced understanding of heavy hadron dynamics and precision measurements. Given the notable enhancements in predictive accuracy, we strongly recommend the utilisation of \textsf{Herwig~7} in future high-precision experimental analyses. We are pleased to announce that these improvements will be made publicly available in the forthcoming \textsf{Herwig-7.3.0} release, underscoring its potential as a reliable computational tool for the broader high-energy physics community.

\section*{Acknowledgements}
\noindent We thank our fellow Herwig authors for useful discussions. This work has received funding from the European Union’s Horizon 2020 research and innovation program as part of the Marie Skłodowska-Curie Innovative Training Network MCnetITN3 (grant agreement no.~722104). \textit{MRM} is also supported by the UK Science and Technology Facilities Council (grant numbers~ST/P001246/1). 

\appendix

\section{Signatures of Heavy Hadrons in the $\Upsilon(4S)$ Continuum Region}
\label{sec:AppDecay}

\setcounter{table}{0}
\renewcommand{\thetable}{A\arabic{table}} 

We have compiled all existing experimental data on heavy mesons and baryons in the $\Upsilon(4S)$ ``continuum region"~\cite{Albrecht:1988dj,Albrecht:1989pa,Albrecht:1989yi,Albrecht:1991ss,Alexander:1993nq,Albrecht:1992zh,Albrecht:1995qx,Avery:1994yc,Bergfeld:1994af,Artuso:2004pj,Seuster:2005tr,Aubert:2002ue,Briere:2000np,Aubert:2006bk,Albrecht:1990zk,Albrecht:1997qa,Albrecht:1993pt,Albrecht:1988gc,Alexander:1999ud,Aubert:2005cu,Aubert:2006cp,Aubert:2006rv,Aubert:2006je,Aubert:2007bt,Avery:1990bc,Avery:1995ps,Bowcock:1989qh,Brandenburg:1996jc,Edwards:1994ar,Gibbons:1996yv,Jessop:1998wt,Mizuk:2004yu,Niiyama:2017wpp,Aubert:2001pd,Abe:2001za}, where $B$ mesons are produced directly from $e^+e^-$ annihilation, bypassing the formation of an $\Upsilon(4S)$ state. This is distinct from the ``resonance region", where $e^+e^-$ collisions are more likely to form an $\Upsilon(4S)$ state that then decays into $B$ mesons. This data was utilized to study the effects of employing the HQET and the spin-flavour symmetry on the overall balance of heavy hadron production rate in \textsf{Herwig-7.3.0} and have been particularly beneficial in our general tuning efforts. 

The tables that follow offer a detailed look at various measurements. Specifically, Tables~\ref{tab9} and \ref{tab10} display results for non-strange and strange heavy charmed mesons, respectively. Tables~\ref{tab11} and \ref{tab12} focus on the experimental data for non-strange and strange charm baryon production in the $\Upsilon$ energy region. Lastly, Table~\ref{tab13} provides insights into the observations of charmonium states in the continuum region.

\begin{table}[H]
\centering
\resizebox{0.65\textwidth}{!}{%
\begin{tabular}{|c||c c c l|}
    \hline
    Hadron & Energy/GeV & Experiment & Observable & Notes \\
    \hline \hline
\multirow{6}{*}{$D^0$} 
    & 10.47 & ARGUS \cite{Albrecht:1991ss} & $\sigma$                   & \\
    &       &                             & $1/N\text{d}N/\text{d}x_p$ & \\
    & 10.52 & BELLE  \cite{Seuster:2005tr} & $\sigma$                   & \\
    & & & $\text{d}\sigma/\text{d}x_p$ & \\
    & 10.5 & CLEO \cite{Artuso:2004pj} & $\sigma$  & \\
    & & & $1/N\text{d}N/\text{d}x_p$ & \\
    \hline
\multirow{6}{*}{$D^+$} 
    & 10.47 & ARGUS \cite{Albrecht:1991ss} & $\sigma$  & \\
    & & & $1/N\text{d}N/\text{d}x_p$ & \\
    & 10.52 & BELLE  \cite{Seuster:2005tr} & $\sigma$  & \\
    & & & $\text{d}\sigma/\text{d}x_p$ & \\
    & 10.5 & CLEO \cite{Artuso:2004pj} & $\sigma$  & \\
    & & & $\text{d}\sigma/\text{d}x_p$ & \\
    \hline
\multirow{4}{*}{$D^{*0}$} 
    & 10.52 & BELLE \cite{Seuster:2005tr} & $\sigma$  & \\
    & & & $\text{d}\sigma/\text{d}x_p$ & \\
    & 10.5 & CLEO \cite{Artuso:2004pj} & $\sigma$  & \\
    & & & $\text{d}\sigma/\text{d}x_p$ & \\
    \hline
\multirow{6}{*}{$D^{*+}$} 
    & 10.47 & ARGUS \cite{Albrecht:1991ss} & $\sigma$  & \\
    & & & $1/N\text{d}N/\text{d}x_p$ & \\
    & 10.52 & BELLE \cite{Seuster:2005tr} & $\sigma$  & \\
    & & & $\text{d}\sigma/\text{d}x_p$ & \\
    & 10.5 & CLEO \cite{Artuso:2004pj} & $\sigma$  & \\
    & & & $\text{d}\sigma/\text{d}x_p$ & \\
    \hline
\multirow{2}{*}{$D_1(2420)^0$} 
    & 10.58 & ARGUS \cite{Albrecht:1989pa} & $\sigma\times\text{Br}$  & \\
    & & & $1/N\text{d}N/\text{d}x_p$ & for $x_p>0.55$ \\
    \hline
\multirow{2}{*}{$D_1(2420)^+$} 
    & 10.58 & CLEO \cite{Bergfeld:1994af} & $\sigma\times\text{Br}$  & \\
    & & & $1/N\text{d}N/\text{d}x_p$ & for $x_p>0.5$ \\
    \hline
\multirow{5}{*}{$D^{\star}_2(2460)^0$} 
    & 10.58 & ARGUS \cite{Albrecht:1988dj} & $\sigma\times\text{Br}$  & \\
    & 10.58 & ARGUS \cite{Albrecht:1989pa} & $\sigma\times\text{Br}$  & \\
    & 10.58 & CLEO \cite{Avery:1994yc} & $\sigma\times\text{Br}$  & \\
    & & & $1/N\text{d}N/\text{d}x_p$ & for $x_p>0.5$ \\
    & & & $1/N\text{d}N/\text{d}x_p$ & for $x_p>0.55$ \\
    \hline
\multirow{2}{*}{$D^{\star}_2(2460)^+$} 
    & 10.58 & CLEO \cite{Bergfeld:1994af} & $\sigma\times\text{Br}$  & \\
    & & & $1/N\text{d}N/\text{d}x_p$ & for $x_p>0.5$ \\
    \hline
\end{tabular}}
\caption{Measurements of non-strange charm meson production in the $\Upsilon$ energy region.}
\label{tab9}
\end{table}

\begin{table}[H]
\centering
\resizebox{0.7\textwidth}{!}{%
\begin{tabular}{|c||c c c l|}
    \hline
    Hadron & Energy/GeV & Experiment & Observable & Notes \\
    \hline \hline
\multirow{5}{*}{$D_s^+$} & 10.58 & BaBar \cite{Aubert:2002ue} & $\sigma$  & \\
    & 10.52 & BELLE \cite{Seuster:2005tr} & $\sigma$  & \\
    & & & $\text{d}\sigma/\text{d}x_p$ & \\
    & & & $1/N\text{d}N/\text{d}x_p$ & \\
    & 10.5 & CLEO \cite{Briere:2000np} & $\text{Br}\times\text{d}\sigma/\text{d}x_p$ & \\
    \hline
\multirow{3}{*}{$D_s^{*+}$} & 10.58 & BaBar \cite{Aubert:2002ue} & $\sigma$  & \\
    & & & $1/N\text{d}N/\text{d}x_p$ & \\
    & 10.5 & CLEO \cite{Briere:2000np} & $\text{Br}\times\text{d}\sigma/\text{d}x_p$ & \\
    \hline
\multirow{2}{*}{$D_{s2}^+$} & 10.58 & ARGUS \cite{Albrecht:1995qx} & $\sigma\times\text{Br}$ & for $x_p>0.7$ + extrapolation\\
    & & & $1/N\text{d}N/\text{d}x_p$ & for $x_p>0.6$ \\
    \hline
\multirow{2}{*}{$D_{s0}^{\star}(2317)^+$} & 10.6 & BaBar \cite{Aubert:2006bk} & $\sigma\times\text{Br}$ & for $p*>3.2$ GeV\\
    & & & $1/N\text{d}N/\text{d}p^{\star}$ & for $p*>3.2$ GeV\\
    \hline
\multirow{2}{*}{$D_{s1}^{\star}(2460)^+$} & 10.6 & BaBar \cite{Aubert:2006bk} & $\sigma\times\text{Br}$ & for $p*>3.2$ GeV\\
    & & & $1/N\text{d}N/\text{d}p^{\star}$ & for $p*>3.2$ GeV\\
    \hline
\multirow{3}{*}{$D_{s1}^{\star}(2536)^+$} & 10.6 & ARGUS \cite{Albrecht:1989yi} & $\sigma\times\text{Br}$ &\\
    & & & $1/N\text{d}N/\text{d}x_p$ & for $x_p>0.5$\\
    & 10.4 & ARGUS \cite{Albrecht:1992zh} & $\sigma\times\text{Br}$ &\\
    \hline
\multirow{2}{*}{$D_{s1}^{\star}(2536)^+$} & 10.6 & BaBar \cite{Aubert:2006bk} & $\sigma\times\text{Br}$ & for $p*>3.2$ GeV\\
    & 10.6 & CLEO \cite{Alexander:1993nq} & $\sigma\times\text{Br}$ &\\
    & & & $1/N\text{d}N/\text{d}x_p$ & for $x_p>0.5$\\
    \hline
  \end{tabular}}
  \caption{Measurements of strange charm meson production in the $\Upsilon$ energy region.}
  \label{tab10}
\end{table}

\begin{table}[H]
\centering
\resizebox{0.7\textwidth}{!}{%
  \begin{tabular}{|c||c c c l|}
    \hline
    Hadron & Energy/GeV & Experiment & Observable & Notes \\
    \hline \hline
\multirow{8}{*}{$\Lambda_c^+$} & 10.54 & BaBar \cite{Aubert:2006cp} &
    $\sigma$ & \\
    & & & $1/N\text{d}N/\text{d}x_p$ & \\
    & 10.52 & BELLE \cite{Seuster:2005tr} &
    $\sigma$  & \\
    & & & $\text{d}\sigma/\text{d}x_p$ & \\
    & 10.52 & BELLE \cite{Niiyama:2017wpp} & $\sigma$ & \\
    & & & $\text{d}\sigma/\text{d}x_p$ & \\
    & 10.5 & CLEO \cite{Avery:1990bc} & $\text{Br}\times\text{d}\sigma/\text{d}x_p$ & \\
    \hline
\multirow{5}{*}{$\Lambda_c(2595)^+$} & 10.4 & ARGUS \cite{Albrecht:1997qa} &
    $\sigma\times\text{Br}$ & for $x_p>0.7$ + extrapolation\\
    & & & $1/N\text{d}N/\text{d}x_p$ & for $x_p>0.5$ \\
    & 10.6& CLEO \cite{Edwards:1994ar} & $\sigma/\sigma(\Lambda_c^+)$ & \\
    & & & $1/N\text{d}N/\text{d}x_p$ & for $x_p>0.5$ \\
    & 10.52 & BELLE \cite{Niiyama:2017wpp} & $\sigma$ & \\
    & & & $\text{d}\sigma/\text{d}x_p$ & \\
    \hline
\multirow{4}{*}{$\Lambda_c(2625)^+$} & 10.4 & ARGUS \cite{Albrecht:1993pt} &
    $\sigma\times\text{Br}$ & extrapolation to whole region\\
    & & & $1/N\text{d}N/\text{d}x_p$ & for $x_p>0.5$ \\
    & 10.6& CLEO \cite{Edwards:1994ar} &
    $\sigma/\sigma(\Lambda_c^+)$ & \\
    & & & $1/N\text{d}N/\text{d}x_p$ & for $x_p>0.5$ \\
    \hline \hline
\multirow{4}{*}{$\Sigma_c^0$} & 10.58  & ARGUS \cite{Albrecht:1988gc} & $\sigma/\sigma(\Lambda_c^+)$ & \\
    &       &                             & $1/N\text{d}N/\text{d}x_p$ & averaged over $\Sigma_c^{0,++}$ \\
    & 10.52 & BELLE \cite{Niiyama:2017wpp} & $\sigma$ & \\
    &       &                             & $\text{d}\sigma/\text{d}x_p$ & \\
    \hline
\multirow{3}{*}{$\Sigma_c^{++}$} & 10.58  & ARGUS \cite{Albrecht:1988gc} & $\sigma/\sigma(\Lambda_c^+)$ & averaged over $\Sigma_c^{0,++}$ \\
    &       &                             & $1/N\text{d}N/\text{d}x_p$ & averaged over $\Sigma_c^{0,++}$ \\
    & 10.58 & CLEO \cite{Bowcock:1989qh}  & $\sigma/\sigma(\Lambda_c^+)$ & averaged over $\Sigma_c^{0,++}$ \\
    \hline
\multirow{2}{*}{$\Sigma_c^{*0}$} & 10.52 & BELLE \cite{Niiyama:2017wpp} & $\sigma$ & \\
    & & & $\text{d}\sigma/\text{d}x_p$ & \\
    \hline
\multirow{2}{*}{$\Sigma_c^{*++}$} & 10.5& CLEO \cite{Brandenburg:1996jc} &$\sigma/\sigma(\Lambda_c^+)$ & averaged over $\Sigma^{*0,++}_c$ \\
    &     &    & $\text{d}\sigma/\text{d}x_p$ & $x_p>0.5$ averaged over $\Sigma^{*0,++}_c$\\
    \hline
\multirow{2}{*}{$\Sigma_c(2800)$} & 10.52 & BELLE \cite{Mizuk:2004yu} &
    $\sigma\times\text{Br}$ & extrapolation to whole region\\
    & & & $1/N\text{d}N/\text{d}x_p$ & for $x_p>0.5$ \\
    \hline
  \end{tabular}}
  \caption{Measurements of non-strange charm baryon production in the $\Upsilon$ energy region.}
  \label{tab11}
\end{table}

\begin{table}[H]
\centering
\resizebox{0.8\textwidth}{!}{%
  \begin{tabular}{|c||c c c l|}
    \hline
    Hadron & Energy/GeV & Experiment & Observable & Notes \\
    \hline \hline
    \multirow{2}{*}{$\Xi_c^0$} & 10.5 & ARGUS \cite{Albrecht:1990zk}  &
    $\sigma\times\text{Br}$ & for $x_p>0.5$ + extrapolation\\
    & & & $1/N\text{d}N/\text{d}x_p$ & for $x_p>0.55$ averaged over $\Xi_c^{0,+}$ \\
    \hline
    \multirow{4}{*}{$\Xi_c^0$} & 10.58 & BaBar \cite{Aubert:2005cu} &
    $\sigma\times\text{Br}$ & \\
    & & & $\text{d}\sigma/\text{d}p\times\text{Br}$ & \\
    & 10.52 & BELLE \cite{Niiyama:2017wpp} & $\sigma\times\text{Br}$ & \\
    & & & $\text{d}\sigma/\text{d}x_p\times\text{Br}$ & \\
    \hline
    \multirow{2}{*}{$\Xi_c^+$} & 10.5 & ARGUS \cite{Albrecht:1990zk}  &
    $\sigma\times\text{Br}$ & for $x_p>0.5$ + extrapolation\\
    & & & $1/N\text{d}N/\text{d}x_p$ & for $x_p>0.55$ averaged over $\Xi_c^{0,+}$ \\
    \hline
    \multirow{4}{*}{$\Xi_c^{\prime0}$} & 10.5 & BaBar \cite{Aubert:2006rv} &
    $\sigma\times\text{Br}$ &  Br is for the $\Xi_c^0$ decay \\
    & & & $1/N\text{d}N/\text{d}x_p$ & also includes $\Upsilon(4S)$ decays \\
    & & CLEO \cite{Jessop:1998wt} & $\sigma/\sigma(\Xi_c^0)$ & \\
    & & & $1/N\text{d}N/\text{d}x_p$ & for $x_p>0.5$ averaged over $\Xi_c^{\prime0,+}$ \\
    \hline
    \multirow{4}{*}{$\Xi_c^{\prime+}$} & 10.5 & BaBar \cite{Aubert:2006rv} &
    $\sigma\times\text{Br}$ &  Br is for the $\Xi_c^+$ decay \\
    & & & $1/N\text{d}N/\text{d}x_p$ & also includes $\Upsilon(4S)$ decays \\
    & 10.5 & CLEO \cite{Jessop:1998wt} & $\sigma/\sigma(\Xi_c^+)$ & \\
    & & & $1/N\text{d}N/\text{d}x_p$ & for $x_p>0.5$ averaged over $\Xi_c^{\prime0,+}$ \\
    \hline
    \multirow{2}{*}{$\Xi_c^{*+}$} & 10.6 & CLEO \cite{Gibbons:1996yv} &
    $\sigma\times\text{Br}/\sigma(\Xi_c^+)$ & \\
    & & & $1/N\text{d}N/\text{d}x_p$ & for $x_p>0.5$ \\
    \hline
    \multirow{2}{*}{$\Xi_c^{*0}$} & 10.6 & CLEO \cite{Avery:1995ps} &
    $\sigma\times\text{Br}/\sigma(\Xi_c^+)$ & \\
    & & & $1/N\text{d}N/\text{d}x_p$ & for $x_p>0.5$ \\
    \hline
    \multirow{1}{*}{$\Xi_{c1}^+$} & 10.6 & CLEO \cite{Alexander:1999ud} & $1/N\text{d}N/\text{d}x_p$ & for $x_p>0.5$ \\
    \hline
    \multirow{1}{*}{$\Xi_{c1}^0$} & 10.6 & CLEO \cite{Alexander:1999ud} & $1/N\text{d}N/\text{d}x_p$ & for $x_p>0.5$ \\
    \hline \hline
    \multirow{4}{*}{$\Omega_c^0$}  & 10.52 & BELLE \cite{Niiyama:2017wpp} & $\sigma\times\text{Br}$ & \\
    & & & $\text{d}\sigma/\text{d}x_p\times\text{Br}$ & \\
    & 10.5 & BaBar \cite{Aubert:2007bt} & $\sigma\times\text{Br}$ & \\
    & & & $1/N\text{d}N/\text{d}x_p$ & also includes $\Upsilon(4S)$ decays \\
    \hline
    \multirow{1}{*}{$\Omega_c^{*0}$} & 10.5 & BaBar \cite{Aubert:2006je} & $\sigma/\sigma(\Omega_c)$ & $x_p>0.5$ \\
    \hline
  \end{tabular}}
  \caption{Measurements of strange charm baryon production in the $\Upsilon$ energy region.}
  \label{tab12}
\end{table}

\begin{table}[H]
\centering
\resizebox{0.45\textwidth}{!}{%
  \begin{tabular}{|c||c c c|}
    \hline
    Hadron & Energy/GeV & Experiment & Observable \\
    \hline
    \multirow{3}{*}{$J/\psi$} & 10.57 & BaBar \cite{Aubert:2001pd} & $\sigma$ \\
    & 10.6  & BELLE \cite{Abe:2001za} & $\sigma$ \\
    & & & $1/N\text{d}N/\text{d}x_p$ \\
    \hline
    \multirow{2}{*}{$\psi(2S)$} & 10.6  & BELLE \cite{Abe:2001za} & $\sigma$ \\
    & & & $1/N\text{d}N/\text{d}x_p$ \\
    \hline
  \end{tabular}}
  \caption{Measurements of charmonium production in the $\Upsilon$ energy region.}
  \label{tab13}
\end{table}

\end{document}